\documentclass[reqno,11pt]{amsart}
\usepackage[utf8]{inputenc}
\usepackage{graphicx}
\usepackage{slashed}
\usepackage{amscd}
\usepackage{amssymb}
\usepackage{esint}
\usepackage{enumitem}
\usepackage{comment}
\usepackage{amsmath}
\usepackage{dsfont}
\usepackage[mathscr]{eucal}
\usepackage{pstricks}

\numberwithin{equation}{section}
\allowdisplaybreaks[1]

\title[The Homogeneous Causal Action Principle]{The Homogeneous Causal Action Principle \\
on a Compact Domain in Momentum Space}

\author[F.\ Finster]{Felix Finster}
\author[M.\ Frankl]{Michelle Frankl}
\author[C.\ Langer]{Christoph Langer \\ \\ May 2022}
\address{Fakult\"at f\"ur Mathematik \\ Universit\"at Regensburg \\ D-93040 Regensburg \\ Germany}
\email{finster@ur.de, michelle.frankl@stud.ur.de, Christoph.Langer@math.ur.de}

\newtheorem{Def}{Definition}[section]
\newtheorem{Thm}[Def]{Theorem}
\newtheorem{Prp}[Def]{Proposition}
\newtheorem{Lemma}[Def]{Lemma}
\newtheorem{Remark}[Def]{Remark}
\newtheorem{Corollary}[Def]{Corollary}
\newtheorem{Example}[Def]{Example}

\newcommand{\Thanks}{\vspace*{.5em} \noindent \thanks}
\newcommand{\beq}{\begin{equation}}
\newcommand{\eeq}{\end{equation}}
\newcommand{\Proof}{\begin{proof}}
	\newcommand{\QED}{\end{proof} \noindent}
\newcommand{\QEDrem}{\ \hfill $\Diamond$}

\newcommand{\la}{\langle}
\newcommand{\ra}{\rangle}

\newcommand{\Sl}{\mbox{$\prec \!\!$ \nolinebreak}}
\newcommand{\Sr}{\mbox{\nolinebreak $\succ$}}

\newcommand{\C}{\mathbb{C}}
\newcommand{\R}{\mathbb{R}}
\newcommand{\1}{\mbox{\rm 1 \hspace{-1.05 em} 1}}

\newcommand{\N}{\mathbb{N}}

\newcommand{\Tr}{{\mathrm{Tr}}}
\renewcommand{\L}{{\mathcal{L}}}
\newcommand{\Sact}{{\mathcal{S}}}
\newcommand{\T}{{\mathcal{T}}}

\newcommand\Q{{\mathcal{Q}}}

\renewcommand{\H}{\mathscr{H}}

\newcommand{\Lin}{\text{\rm{L}}}
\newcommand{\F}{{\mathscr{F}}}

\newcommand{\Symm}{\text{\rm{Symm}}}

\DeclareMathOperator{\re}{Re}

\DeclareMathOperator{\supp}{supp}

\newcommand{\scrM}{\mycal M}
\newcommand{\hscrM}{\,\,\hat{\!\!\mycal M}}

\newcommand{\itemD}{\item[{\raisebox{0.125em}{\tiny $\blacktriangleright$}}]}

\newcommand{\bitem}{\begin{itemize}[leftmargin=2em]}
\newcommand{\eitem}{\end{itemize}}

\newcommand{\G}{{\mathscr{G}}}

\DeclareFontFamily{OT1}{rsfso}{}
\DeclareFontShape{OT1}{rsfso}{m}{n}{ <-7> rsfso5 <7-10> rsfso7 <10-> rsfso10}{}
\DeclareMathAlphabet{\mycal}{OT1}{rsfso}{m}{n}

\begin{document}
\maketitle

\begin{abstract}
The homogeneous causal action principle on a compact domain
of momentum space is introduced. The connection to causal fermion systems is worked out.
Existence and compactness results are reviewed. The Euler-Lagrange equations are derived and analyzed
under suitable regularity assumptions.
\end{abstract}
\tableofcontents

\thispagestyle{empty}

\section{Introduction} \label{secintro}
The theory of {\em{causal fermion systems}} is a recent approach to fundamental physics
(see the reviews~\cite{review, dice2014}, the introduction~\cite{langerintro},
the textbooks~\cite{cfs, intro} or the website~\cite{cfsweblink}).
In this approach, spacetime and all objects therein are described by a measure~$\rho$
on a set~$\F$ of linear operators on a Hilbert space~$(\H, \la .|. \ra_\H)$. 
The physical equations are formulated by means of the so-called {\em{causal action principle}},
a nonlinear variational principle where an action~$\Sact$ is minimized under variations of the measure~$\rho$.
The {\em{homogeneous}} causal action principle was introduced in~2008 in~\cite[Section~4]{continuum}
as a formulation of the causal action principle for systems which are invariant under
translations in space and time. 
Such translation invariant causal fermion systems are of interest because they provide candidates
for vacuum spacetimes involving a regularization on a microscopic scale,
typically associated to the Planck length~$\ell_P \approx 10^{-35}$ meters.
Therefore, the study of the homogeneous causal action principle
should give insight into the structure of physical spacetime on the Planck scale.

In the past years, the theory of causal fermion systems was developed further, and the
mathematical setting evolved. Therefore, we take the present article as an opportunity
for giving a coherent introduction to the homogeneous causal action principle, which makes the
connection to the modern formulation and terminology and gives an up-to date review of the present status
of this nonlinear variational principle.
Moreover, our main concern is the derivation of the {\em{Euler-Lagrange equations}}.
Here we face the main difficulty that the homogeneous causal action principle
involves nonlinear constraints (both equality and inequality constraints) which cannot be
treated with the standard Lagrange multiplier method.
The results and methods underlying the present paper are based on
more detailed expositions in~\cite{langermaster, langerdr, frankl}.

In order to motivate the problem starting from the physical applications,
we first note that the causal fermion system
describing the vacuum in Minkowski space may be described by the {\em{unregularized
kernel of the fermionic projector}}~$P(x,y)$, being a tempered distribution defined as a Fourier
transform,
\beq \label{Punreg}
P(x,y) = \int_{\R^4} \frac{d^4p}{(2 \pi)^4}\: (\slashed{p}+m)\: \delta(p^2-m^2)\: \Theta(-p^0)\: e^{-ik(x-y)}
\eeq
(here~$\slashed{p}=\gamma^j p_j$ denotes the contraction with the Dirac matrices, $\Theta$ is the
Heaviside function, and $k(x-y)$ denotes the Minkowski inner product of signature~$(+,-,-,-)$;
for the physical background and the notation see for example the textbooks~\cite{cfs, intro}).
In order to make mathematical sense of the causal action principle, one needs to introduce
a {\em{regularization}} (for details and the underlying physical concepts
see for example~\cite[Section~1.2]{cfs}).  As a simple example,
one may regularize by
inserting a convergence-generating factor~$e^{\varepsilon p^0}$ into the integrand in~\eqref{Punreg},
which mollifies the kernel on a microscopic length scale~$\varepsilon$ (which can be thought of as the Planck length).
The basic concept behind causal fermion systems is that
the regularized objects are the physical objects. In particular, the regularized kernel of the fermionic
projector describes the structure of spacetime on microscopic scales where Minkowski space
is no longer the appropriate mathematical model. With this in mind, it is an important
task to understand how the regularization looks like for minimizers of the causal action principle.
In order to address this problem in the simplest possible setting, in~\cite[Section~4]{continuum}
an a-priori momentum cutoff was introduced. To this end, one restricts attention to
a compact subset~$\hat{K}$ of momentum space which is chosen at the very beginning. 
Thus, instead of~\eqref{Punreg}, one considers the kernel
\beq \label{Preg}
P^\text{reg}(x,y) = \int_{\hat{K}} \frac{d^4p}{(2 \pi)^4}\: \hat{P}(p)\: e^{-ik(x-y)}
\eeq
involving an arbitrary matrix-valued distribution~$\hat{P}(p)$.
By suitably varying~$\hat{P}(p)$ one gets a well-posed variational principle,
meaning that a minimizer~$\hat{P}(p)$ exists (for more details see below).
It is conjectured that in a suitable limit~$\hat{K} \nearrow \R^4$ when~$\hat{K}$ exhausts~$\R^4$,
the corresponding sequence of minimizers~$\hat{P}(p)$ should go over to the integrand in~\eqref{Punreg}.
This conjecture is underpinned by the {\em{so-called continuum limit analysis}} as carried out in~\cite{cfs},
which shows that, expanding asymptotically for small values of the regularization length~$\varepsilon$ and disregarding certain contributions of higher order in~$\varepsilon$, the unregularized kernel~\eqref{Punreg} indeed arises as a minimizer of the causal action principle.
Proving the above conjecture without using the formalism of the continuum limit
is a challenging open problem, which we will not address here
(see however the discussion in the outlook in Section~\ref{secoutlook}).
Instead, as a preparation for the limit~$\hat{K} \nearrow \R^4$,
the main part of the present paper is devoted to analyzing the variational principle
{\em{on a compact domain}}~$\hat{K}$
of momentum space. Our main goal is to clarify the structure of the resulting minimizers
by deriving the corresponding Euler-Lagrange equations.

Before stating our results, we need to make the setting mathematically precise.
To this end, it is useful to combine the factor~$\hat{P}(p)$ in~\eqref{Punreg}
with the integration measure of the Fourier integral
to an object which we shall give a precise mathematical meaning as a measure~$\nu$
in momentum space, i.e.\ symbolically
\beq \label{nuintro}
d\nu(p) := -\hat{P}(p)\: \frac{d^4p}{(2 \pi)^4} \:.
\eeq
In the example of the unregularized kernel~\eqref{Punreg}, this measure takes the form
\beq \label{nuunreg}
d\nu(p) = -(\slashed{p}+m)\: \delta(p^2-m^2)\: \Theta(-p^0)\: \frac{d^4p}{(2 \pi)^4}\:,
\eeq
where the right side denotes a measure which is singular with respect to the Lebesgue measure,
with support on the lower mass shell.
Moreover, in view of the Dirac matrices, this measure is matrix-valued.
The matrix~$\slashed{p}$ is not Hermitian, but it is symmetric with respect to the
indefinite inner product on spinors, which is often denoted
by~$\overline{\psi} \phi$ with the adjoint spinor~$\overline{\psi}:=\psi^\dagger \gamma^0$
(and the dagger denotes complex conjugation and transposition). Here we denote the
corresponding indefinite inner product space abstractly by~$(V, \Sl .|. \Sr)$; it 
is a four-dimensional complex inner product space of signature~$(2,2)$.
By ``symmetry'' of the operator~$\slashed{p}$ we mean that
\[ \Sl \slashed{p} \psi | \phi \Sr = \Sl \psi | \slashed{p} \phi \Sr \qquad \text{for all~$\psi, \phi \in V$}\:. \]
The matrix in~\eqref{nuunreg} has the additional property that it is positive semi-definite in the sense
that
\[ \Sl \psi \,|\, \big( -(\slashed{p}+m) \big) \psi \Sr \geq 0 \qquad \text{for all~$\psi \in V$} \]
(and all~$p$ in the support of~$\nu$).
It turns out that this positivity property is intimately related to the Hilbert space structure
of the solution space of the Dirac equation. It is also crucial for getting into the setting of causal
fermion systems (for details see Section~\ref{seccfs} below) and for getting a well-posed variational problem
(as was first noted in~\cite{discrete}). For the more general measure~$\nu$ in~\eqref{nuintro}, this
positivity property is incorporated by the concept of a {\em{positive definite measure}} on~$\hat{K}$
with values in~$\Lin(V)$ (see Definition~\ref{defposV}).

Working with such positive definite measures~$\nu$ on~$\hat{K}$, one can adapt the
general causal action principle as introduced for example in~\cite[Section~1.1]{cfs}
to homogeneous kernels of the form~\eqref{Preg} and~\eqref{nuintro}.
The resulting {\em{homogeneous causal action principle}} is introduced in Section~\ref{sechomcap}.
In this variational principle, one minimizes the {\em{causal action}}~\ref{Sdef} under variations of~$\nu$
in the class of positive definite regular Borel measures on~$\hat{K}$,
under the {\em{trace constraint}}~\eqref{TC} and the {\em{dimension constraint}}~\eqref{DC}.
After reviewing compactness and existence results, our main concern is the derivation of the
corresponding Euler-Lagrange (EL) equations. These are variational inequalities of a
novel type, as we now briefly summarize.
Under suitable regularity assumptions,
the first variation of the causal action can be written as (for details see Proposition~\ref{prpvariation})
\[  \delta \Sact =  2 \int_{\hat{K}} \Tr \Big( \hat{Q}(p)\: d \big(\delta \nu \big) (p)\Big) \]
with a matrix-valued kernel~$\hat{Q} \in C^0(\hat{K}, \Lin(V))$.
The EL equations as formulated in Theorem~\ref{thmEL}
state that, for suitable real parameters~$\alpha$ and~$\beta$
(which can be thought of as the Lagrange parameters corresponding to the
trace constraint and the dimension constraint, respectively) and a suitable
signature operator~$S$ (see~\eqref{Sigdef} and the remark after Theorem~\ref{thmexist}), the operator
\[ \hat{Q}(p) - \alpha \,\1- \beta\, S \quad \text{is positive semi-definite on~$(V, \Sl .|. \Sr)$} \]
for all~$p \in \hat{K}$.
Moreover, this operator has a non-trivial kernel on the support of the measure~$\nu$, and the measure~$\nu$
vanishes except on this kernel in the sense that
\[ \big( \hat{Q}(p) -\alpha \,\1 - \beta\, S \big) \:d\nu(p) = 0 = 
d\nu(p) \:\big( \hat{Q}(p) -\alpha \,\1 - \beta\, S \big) \]
(here the product denotes the multiplication of a function with a measure;
for example, $\hat{Q}(p) \:d\nu(p)$ could be written in a shorter form as~$\hat{Q}\, \nu$).
These EL equations, which are formulated in terms of positivity properties of linear operators
and the vanishing of certain operator products, have a novel structure which is quite different
from that for classical variational principles or for the general causal action principle for causal fermion systems
(cf.\ for example~\cite[Chapter~7]{intro}).
The form of our EL equations reflects the specific structure of the homogeneous causal action principle.
Our derivation requires new mathematical methods which will be developed
in Sections~\ref{secpush} and~\ref{secEL}.

The paper is organized as follows. After introducing the homogeneous action principle
(Section~\ref{sechomcap}), the connection to causal fermion systems is worked out (Section~\ref{seccfs}).
In Section~\ref{secdecomp} it is shown how a positive definite measure~$\nu$ can be
decomposed into what we refer to as the {\em{sea measure}} and the {\em{particle measure}}.
Section~\ref{secexist} is devoted to an outline of compactness and existence results;
the existence result relevant for our purposes is stated in Theorem~\ref{thmexist}.
In Section~\ref{secfirstvary} first variations of the causal action are computed.
As a simplification, in Section~\ref{secpush} the first variations and the constraints are formulated
with a positive definite measure on the set~$\Symm(V)$ of symmetric linear operators on
the indefinite inner product space~$(V, \Sl .|. \Sr)$.
In Section~\ref{secEL} the Euler-Lagrange equations are worked out.
The main difficulty is to preserve the trace constraint and the dimension constraint
in the variation; this is achieved by a suitable transformation of the positive semi-definite
operators on~$V$ (see~\eqref{s12vary}).
Having treated the constraints, variations which change the support of~$\nu$ give rise to variational
inequalities (Subsection~\ref{secchange}), whereas variations preserving the support 
yield variational equations (Subsection~\ref{secfix}).
Combining these results with a variant of the Cauchy-Schwarz inequality for positive semi-definite
operators on indefinite inner product spaces (Lemma~\ref{lemmacsu}),
we obtain the general form of our EL equations (Theorem~\ref{thmEL}).
In Section~\ref{secpointwise} we analyze a variational principle for linear operators on~$V$
which clarifies the role of the Lagrange parameters~$\alpha$ and~$\beta$ in our
EL equations. In Section~\ref{secoutlook} we conclude the paper with a brief outlook.

\section{The Homogeneous Causal Action Principle} \label{sechomcap}
The homogeneous causal action principle was introduced in~\cite[Section~4]{continuum}.
We now recall the setting in a slightly modified form which is most convenient for our purposes.
Given a parameter~$n \geq 1$ (the {\em{spin dimension}}),
we let~$(V, \Sl .|. \Sr)$ be a complex indefinite inner product space of dimension~$2n$ and signature~$(n,n)$.
We let~$\scrM$ be a four-dimensional real vector space (``position space'')
and~$\mu$ a translation invariant Borel measure (i.e., the Haar measure
corresponding to the translation group; in a basis, this measure is a positive
constant times the Lebesgue measure on~$\R^4$). 
Moreover, we denote the dual space of~$\scrM$
by~$\hscrM$ (``momentum space''). 
Let~$\hat{K} \subset \hscrM$ be a compact subset of momentum space.
For the sake of simplicity, we change the conventions in~\cite{continuum, langerdr}
and consider {\em{positive}} (instead of negative) definite measures.

\begin{Def} \label{defposV}
Consider a regular Borel measure~$\nu$ on a compact set~$\hat{K} \subset \hscrM$ taking
values in~$\Lin(V)$ with the following properties:
\bitem
\item[{\rm{(i)}}] For every~$v \in V$, the measure~$\Sl v | \nu v \Sr$ is a finite signed measure.
\item[{\rm{(ii)}}] For every Borel set~$\Omega \subset \hat{K}$, the operator~$\nu(\Omega)
\in \Lin(V)$ is positive in the sense that
\[ 
\Sl v \,|\, \nu(\Omega)\, v \Sr \geq 0 \qquad \text{for all $v \in V$}\:. \]
\eitem
Then~$\nu$ is called a {\bf{positive definite measure}} on~$\hat{K}$ with values in~$\Lin(V)$.
\end{Def} \noindent
Positivity of the operator~$\nu(\Omega)$ implies in particular that this operator is symmetric with respect to
the indefinite inner product, i.e.\
\beq \label{nusymm}
\Sl u \,|\, \nu(\Omega) \,v \Sr = \Sl \nu(\Omega) \,u \,|\,  v \Sr \qquad \text{for all~$u, v \in V$}\:.
\eeq
Moreover, the operator has a real spectrum and (counting algebraic multiplicities) at most~$n$ positive and at most~$n$ negative eigenvalues (for details see~\cite[Lemma~4.2]{discrete}).

Given a positive definite measure on~$\hat{K}$, we introduce the {\em{kernel of the fermionic
projector}}~$P(\xi)$ by
\beq \label{Pxidef}
P(\xi) := -\int_{\hat{K}} e^{i p \xi} \:d\nu(p) \:.
\eeq
\begin{Lemma} \label{lemmaC0} The kernel of the fermionic projector is continuous,
\[ P \in C^0 \big( \scrM, \Lin(V) \big) \:. \]
\end{Lemma}
\Proof In view of the polarization formula, it suffices to consider the expectation values~$\Sl v | P(\xi) v \Sr$.
Continuity follows from the estimate
\begin{align*}
\Big| \Sl & v \,|\, P(\xi)\, v \Sr - \Sl v \,|\, P(\xi')\, v \Sr \Big| = \bigg| \int_{\hat{K}} \big( e^{i p \xi} 
- e^{i p \xi'} \big) \:d \Sl v | \nu v \Sr(p) \bigg| \\
&\leq \sup_{p \in \hat{K}} \big| e^{i p \xi}  - e^{i p \xi'} \big| \: \big| \Sl v | \nu v \Sr \big|
\xrightarrow{\xi' \rightarrow \xi} 0 \:,
\end{align*}
where in the last step we used that the exponential converges uniformly for~$p$ in a compact set.
Here~$|.|$ denotes the total variation of a signed measure, which is finite in view of
Definition~\ref{defposV}~(i) (we remark that the signed measure~$\Sl v | \nu v \Sr$ is even positive
by Definition~\ref{defposV}~(ii), so that~$| \Sl v | \nu v \Sr| = \Sl v | \nu(\hat{K}) v \Sr$).
\QED
Moreover, as a consequence of~\eqref{nusymm}, the kernel of the fermionic projector is symmetric in the sense that
\[ P(\xi)^* = P(-\xi) \]
(where the star is the adjoint with respect to the indefinite inner product~$\Sl .|. \Sr$).
We introduce the {\em{closed chain}}~$A(\xi)$ by
\[ A(\xi) := P(\xi)\, P(\xi)^* \]
and denote its eigenvalues (counting algebraic multiplicities) by~$\lambda^\xi_1, \cdots, \lambda^\xi_{2n}$.
We introduce the Lagrangian~$\L$ (which clearly depends on~$\nu$) and the homogeneous causal action by
\begin{align}
\text{\em{Lagrangian:}} && \L(\xi) &= \frac{1}{4n} \sum_{i,j=1}^{2n} \Big( \big|\lambda^\xi_i \big|
- \big|\lambda^\xi_j \big| \Big)^2 \label{Ldef} \\
\text{\em{homogeneous causal action:}} && \Sact(\nu) &= \int_{\scrM} \L(\xi)\: d\mu(\xi) \:. \label{Sdef}
\end{align}
Moreover, we introduce the following constraints. The operator~$\nu(\hat{K})$ has
real eigenvalues (see Lemma~\ref{lemmapositive} below).
Counting multiplicities, we denote these eigenvalues by~$\alpha_1, \ldots, \alpha_{2n}$.
Given real parameters~$c$ and~$f$, we define the following constraints,
\begin{align}
\text{\em{trace constraint:}} && \Tr \big( \nu(\hat{K}) \big) &= \sum_{\ell=1}^{2n} \alpha_\ell = c \label{TC} \\
\text{\em{dimension constraint:}} && \sum_{\ell=1}^{2n} \big|\alpha_\ell\big| &\leq f \:. \label{DC}
\end{align}
The {\em{homogeneous causal action principle}} is to minimize the causal action~\eqref{Sdef}
under variations of~$\nu$ within the class of positive definite regular Borel measures on~$\hat{K}$
with values in~$\Lin(V)$, subject to the constraints~\eqref{TC} and~\eqref{DC}.

Clearly, the constraints~\eqref{TC} and~\eqref{DC} can be fulfilled only if~$|c| \leq f$.
By flipping the signs of both~$\nu$ and~$\la .|. \ra$, we can change the sign of~$c$ arbitrarily.
With this in mind, it is no loss of generality to assume that~$c \geq 0$.
The cases~$c=0$ and~$c=f$ are uninteresting limiting cases. Therefore, we shall assume throughout
this paper that
\beq \label{cfrange}
0 < c < f \:.
\eeq

\section{The Underlying Causal Fermion System} \label{seccfs}
Before entering the analysis of the homogeneous causal action principle, in this section
we clarify the connection to causal fermion systems as defined abstractly as follows
(for details see for example~\cite[Section~1.1]{cfs}).
\begin{Def} \label{defparticle} (causal fermion system) {\em{ 
Given a separable complex Hilbert space~$\H$ with scalar product~$\la .|. \ra_\H$
and a parameter~$n \in \N$ (the {\em{``spin dimension''}}), we let~$\F \subset \Lin(\H)$ be the set of all
self-adjoint operators on~$\H$ of finite rank, which (counting multiplicities) have
at most~$n$ positive and at most~$n$ negative eigenvalues. On~$\F$ we are given
a positive measure~$\rho$ (defined on a $\sigma$-algebra of subsets of~$\F$).
We refer to~$(\H, \F, \rho)$ as a {\em{causal fermion system}}.
}}
\end{Def}
Making this connection precise consists of two steps.
First, we will show how, given a positive definite measure~$\nu$, one can construct
a corresponding causal fermion system. The second step is to show that, starting from a
causal fermion system, one gets a positive definite measure~$\nu$ by a suitable symmetry reduction
(this will be done in Remark~\ref{remsymm} below).
Moreover, we will also explain how the homogeneous causal action principle relates to the
general causal action principle for causal fermion systems.

Let~$\nu$ be a positive definite measure on~$\hat{K}$ with values in~$\Lin(V)$.
We consider continuous and compactly supported test functions on~$\hscrM$ with values in~$V$, 
denoted by
\[ \hat{u}, \hat{v} \in C^0_0(\hscrM, V) \:. \]
On such functions we introduce the positive semi-definite inner product
\beq \label{sprod}
\la .|. \ra_\H \::\: C^0_0(\hscrM, V) \times C^0_0(\hscrM, V) \rightarrow \C \:, \qquad
\la \hat{u} | \hat{v} \ra_\H := \int_{\hat{K}} d \Sl \hat{u}(p) \,|\, \nu(p)\, \hat{v}(p) \Sr \:.
\eeq
Dividing out the null space and taking the completion gives a Hilbert
space~$(\H, \la .|. \ra_\H)$ (details can be found in~\cite{langermaster}).
A Hilbert space vector~$\hat{u} \in C^0_0(\hscrM, V)$ can be represented by a canonical
wave function in~$\scrM$, referred to as the {\em{physical wave function}}. It is obtained
by taking the Fourier transform with respect to the measure~$\nu$,
\beq \label{psiu}
\psi^{\hat{u}}(x) = \int_{\hat{K}} e^{-i p x} \:d\nu(p)\: \hat{u}(p)\:.
\eeq

Now we adapt the standard construction of a causal fermion system from the
physical wave functions using the local correlation operators
(as introduced in~\cite[Section~1]{rrev} and explained in more detail for example in~\cite[Section~1.2]{cfs} or~\cite[Chapter~5]{intro}). 
Note that, since~$\hat{K}$ is compact, the physical wave functions defined by~\eqref{psiu} are continuous
(as one immediately verifies as in the proof of Lemma~\ref{lemmaC0} above).
Therefore, for any~$x \in \scrM$ we may evaluate pointwise and define the sesquilinear form
\[ b_x \::\: C^0_0(\hscrM, V) \times C^0_0(\hscrM, V) \rightarrow \C \:,\qquad
b_x(\hat{u}, \hat{v}) := -\Sl \psi^{\hat{u}}(x) | \psi^{\hat{v}}(x) \Sr \:. \]
\begin{Lemma} The sesquilinear form~$b_x$ is bounded with respect to the scalar product~$\la .|. \ra_\H$, i.e.\
there is a constant~$c>0$ such that
\[ \big| b_x(\hat{u}, \hat{v}) \big| \leq c\: \|\hat{u}\|_\H\: \|\hat{v}\|_\H \qquad \text{for all~$\hat{u}, \hat{v} \in
C^0_0(\hscrM, V)$} \:. \]
\end{Lemma}
\Proof Since the measure~$\nu$ is positive definite, the Cauchy-Schwarz inequality
\[ \bigg| \int_{\hat{K}} d\Sl \hat{f}(p) \,|\, \nu(p)\, \hat{g}(p) \Sr \bigg|
\leq \bigg( \int_{\hat{K}} d\Sl \hat{f}(p) \,|\, \nu(p)\, \hat{f}(p) \Sr \bigg)^\frac{1}{2}
 \bigg( \int_{\hat{K}} d\Sl \hat{g}(p) \,|\, \nu(p)\, \hat{g}(p) \Sr \bigg)^\frac{1}{2} \]
holds. In order to estimate~$\psi^{\hat{u}}$ pointwise in terms of the Hilbert space norm, we
apply this inequality for
\[ \hat{f}(p) = \zeta \in V \qquad \text{and} \qquad \hat{g}(p) = \hat{u}(p)\: e^{-ipx} \:. \]
We thus obtain
\[ \big| \Sl \zeta | \psi^{\hat{u}}(x) \Sr \big|^2 \leq
\Sl \zeta | \nu(\hat{K}) \zeta \Sr \: \int_{\hat{K}} d \Sl \hat{u}(p)\:|\: \nu(p)\: \hat{u}(p) \Sr
= \Sl \zeta | \nu(\hat{K}) \zeta \Sr \: \|\hat{u}\|_\H^2 \:. \]
Choosing a basis of~$V$, this inequality gives an estimate of each component of~$\psi^{\hat{u}}(x)$.
Therefore, there is a constant~$c$ (which depends only on~$n$ and~$\nu(\hat{K})$) such that
for all~$\hat{u}, \hat{v} \in
C^0_0(\hscrM, V)$,
\[ \big| \Sl \psi^{\hat{u}}(x) | \psi^{\hat{v}}(x) \Sr \big| \leq c\: \|\hat{u}\|_\H\: \|\hat{v}\|_\H \:. \]
This concludes the proof.
\QED
This lemma shows that the sesquilinear form~$b_x$ can be continuously extended
to a bounded sesquilinear form
\[ b_x \::\: \H \times \H \rightarrow \C \:. \]
Using the Fr{\'e}chet-Riesz theorem, we can represent this sesquilinear form by a
symmetric linear operator~$F(x)$, which is uniquely defined by the property that
\[ \la \hat{u} \,| F(x)\, \hat{v} \ra_\H = -\Sl \psi^{\hat{u}}(x) | \psi^{\hat{v}}(x) \Sr \;\in\; \F \qquad
\text{for all~$\hat{u}, \hat{v} \in C^0_0(\hscrM, V)$}\:. \]
This operator, referred to as the {\em{local correlation operator}} at~$x$, has at most~$n$
positive and at most~$n$ negative eigenvalues; thus it is an operator in~$\F$.
Varying~$x$, we obtain the {\em{local correlation map}}~$F : \scrM \rightarrow \F$.
Taking the push-forward measure of the volume measure~$d\mu = d^4x$ of~$\scrM$,
\beq \label{rhoconstruct}
\rho := F_* \mu \:,
\eeq
gives 
a causal fermion system~$(\H, \F, \rho)$.

In order to clarify the above structures, it is useful to rewrite the scalar product~\eqref{sprod}
in position space. To this end, we introduce the
{\em{kernel of the fermionic projector}}~$P(x,y)$ for any~$x,y \in \scrM$ by
\beq \label{Pxydef}
P(x,y) := -\int_{\hscrM} e^{i p (y-x)} \:d\nu(p) \;\in\; \Lin(V) \:.
\eeq
Using Plancherel's theorem, the physical wave function~$\psi^{\hat{u}}$ can also be expressed by
\[ 
\psi^{\hat{u}}(x) = -\int_\scrM P(x,y)\: u(y)\: d\mu(y) \:, \]
where~$u$ is the ordinary Fourier transform,
\[ u(x) := \int_{\hscrM} \frac{d^4p}{(2 \pi)^4} \:\hat{u}(p)\: e^{-i p x} \:. \]
Moreover, the scalar product~\eqref{sprod} can be expressed in position space as
\[ 
\la \hat{u} | \hat{v} \ra_\H = -\int_\scrM d^4x \int_\scrM d^4y\; \Sl u(x) \,|\, P(x,y)\, u(y) \Sr \:. \]

We now show how to get from a causal fermion system to a
positive definite measure~$\nu$ by a suitable symmetry reduction
(for further details see again~\cite{langermaster}).
\begin{Remark} {\bf{(homogeneous causal fermion systems)}} \label{remsymm} {\em{
Let~$(\H, \F, \rho)$ be a causal fermion system.
A {\em{symmetry of the causal system}} is a group~$\G$
together with a unitary representation~$U$ on~$\H$ which leaves the measure~$\rho$ invariant, i.e.
\[ \rho\big( U_g\, \Omega\, U_g^{-1} \big) = \rho(\Omega) \qquad
\text{for all~$g \in \G$ and all measurable~$\Omega \subset \F$}\:. \]
Defining {\em{spacetime}} as usual by~$M:= \supp \rho$, it follows that the representation~$U$
leaves~$M$ invariant, i.e.\ $U_g M U_g^{-1} = M$.
Moreover, the induced mapping
\[ T \::\: \G \times M \rightarrow M \:,\qquad (g,x) \mapsto U_g \,x\, U_g^{-1} \]
is an {\em{action}} of~$\G$ on~$M$.

We now restrict attention to a specific group: the group of translations in four-dimensional space,
\[ \G = (\R^4, +) \:. \]
Moreover, we assume that the unitary representation~$U$ of~$\G$ on~$\H$
is {\em{strongly continuous}}. Then, denoting the canonical basis of~$\R^4$ by~$(e_0,\ldots, e_3)$,
the four operators~$U_{e_0}, \ldots, U_{e_3}$ are unitary and mutually commute.
To each one-parameter group~$U_{t e_i}$, we can apply Stone's theorem to obtain a self-adjoint
operator~$A_i$ with $U_{t e_i} = \exp(itA_i)$. Then the resulting operators~$\tanh(A_i)$ are bounded
and mutually commute. The spectral theorem for commuting operators gives a spectral measure~$E$
on~$[-1,1]^4$. The functional calculus yields a corresponding spectral measure on~$\R^4$
for the operators~$A_i$, i.e.\
\[ U_{(t, x, y, z)} = \int_{\R^4} e^{i ( p_0 t + p_1 x + p_2 y + p_3 z )}\: dE_p \:. \]
We write this in the shorter form
\beq \label{Uspec}
U_\xi = \int_{\G^*} e^{i p_j \xi^j}\: dE_p \qquad \text{where} \qquad \xi = (t,x,y,z)\:.
\eeq
Note that, at this stage, we make no use of a scalar product on~$\R^4$.
We can always think of~$\G^*$ as momentum space, the dual space of the translations
in position space~$\G$.

Next, we assume that the {\em{action of~$\G$ on~$M$ is faithful and transitive}}.
Then for every~$x,y \in M$ there is a unique~$\xi \in \G$ such that~$y = T_\xi \,x$.
Fixing~$x$ and varying~$y$, we obtain the identification~$M \simeq \G$. 
Moreover, we have the useful formula
\beq \label{yxid}
y = T_\xi \,x = U_\xi \, x \,U_\xi^{-1} \:.
\eeq
Next, it is convenient to identify all the spin spaces. Recall that for every~$x \in M$, the
corresponding spin space~$(S_x, \Sl .|. \Sr_x)$ is defined by
\[ S_x = x(\H) \:,\qquad \Sl .|. \Sr_x = - \la .| x . \ra_\H\big|_{S_x \times S_x}\:. \]
Thus the mapping~$U_\xi$ gives an isomorphisms of the corresponding spin spaces,
\[ U_\xi \::\: S_x \rightarrow S_y \:. \]
Again fixing~$x$, we obtain the identifications
\[ 
S_y \;\simeq\; S := S_x \qquad \text{given by} \qquad S_y = U_\xi \,S_x \:. \]
Using these identifications, we can simplify the formulas for the fermionic projector. 
Namely,
\begin{align*}
P(x,y) &:= \pi_x y \::\: S_y \rightarrow S_x \\
&\,\,= \pi_x y \,U_{\xi} \::\: S_x \rightarrow S_x \:.
\end{align*}
Using~\eqref{yxid}, 
we obtain
\[ P(x,y) = \pi_x U_{\xi}\, U_{\xi}^{-1}\, y \,U_{\xi} = \pi_x \,U_{\xi}\, x \:. \]
Now we employ the spectral representation~\eqref{Uspec} to obtain
\beq \label{Pxyform}
P(x,y) = \int_{\G^*} e^{i p_j \xi^j}\: \big( \pi_x\, dE_p\, x \big) \:.
\eeq
This formula resembles the Fourier representation~\eqref{Pxydef} of the kernel
of the fermionic projector. In~\cite[Section~4]{continuum}, this formula was used as an
ansatz for homogeneous causal fermion systems.
Now this ansatz has been {\em{derived}} by imposing the action of a symmetry group.

We next prove that the measure~$-\pi_x\, dE_p\, x$ is a positive definite measure
on~$\G^*$ with values in~$\Lin(S_x)$. To this end, for any~$u, v \in S_x \subset \H$ and any measurable set~$V \subset \G^*$,
\begin{align*}
\Sl u \,|\, \big(-\pi_x\, E(V)\, x\big)  v \Sr &= \la u \,|\, x \,\big(\pi_x\, E(V)\, x\big) \, v \ra_\H \\
&= \la u \,|\, x \,\pi_x\, E(V)\, x\, v \ra_\H = \la x u \,|\, E(V)\, x v \ra_\H \:.
\end{align*}
Using that every projection operator is positive semi-definite,
one concludes that this measure is indeed positive definite.

Denoting the measure~$-\pi_x\, dE_p\, x$ by~$d\nu$, the formula~\eqref{Pxyform}
agrees with~\eqref{Pxydef}. This suggests that, taking this measure as the starting
point, the construction leading to~\eqref{rhoconstruct} should give back our homogeneous causal fermion system.
However, proving strong continuity of the resulting group representation is rather subtle.
Sufficient technical assumptions are discussed in~\cite{langermaster}.
}} \QEDrem
\end{Remark}

We finally explain the connection between the homogeneous causal action principle
and the general causal action principle for causal fermion systems
as introduced for example in~\cite[Section~1.1]{cfs}.
The Lagrangian~$\L$ in~\eqref{Ldef} has the same form as in~\cite[Section~1.1]{cfs}
if the vector~$\xi$ is regarded as the difference vector~$\xi=y-x$
(see also~\eqref{Pxydef}). The action~\eqref{Sdef} differs from the general causal action
in that we integrate only over the vector~$\xi$ (instead of over both~$x$ and~$y$).
This difference can be understood immediately from the fact that, for homogeneous systems,
the Lagrangian depends only on~$y-x$. Integrating over both~$x$ and~$y$ would give
an irrelevant, but infinite prefactor. 
Similarly, the trace constraint~\eqref{TC} is obtained from the general trace constraint by
omitting one spacetime integral. The dimension constraint, however, has no correspondence
in the setting of general causal fermion systems. It can be understood as replacing the constraint that
when varying a causal fermion system, the dimension of the Hilbert space~$\H$ is kept fixed.
This is also the motivation for the name {\em{dimension}} constraint
(for more details see Remark~\ref{remdim} below).
With the dimension constraint present, the analog of the boundedness constraint can be left out.
Working with the dimension constraint also has the advantage that, just as the trace constraint,
it is homogeneous of degree one in~$\nu(\hat{K})$. The resulting scaling freedom of minimizers
may simplify the construction of minimizers in non-compact domains of momentum space
(for some more details see Section~\ref{secoutlook}).

\section{Decomposition into the Sea Measure and the Particle Measure} \label{secdecomp}
We now explain that a positive definite measure~$\nu$ on~$\hat{K}$ (see Definition~\ref{defposV})
has a canonical decomposition into a sum of measures which can be interpreted as
describing the particles and anti-particles.
For the construction, we let~$\|.\|_V$ be an arbitrary norm on the indefinite inner product space~$(V, \la .|. \ra)$.
We let~$|\nu|$ be the variation measure of~$\nu$ with respect to this norm, i.e.\
\beq \label{totvar}
|\nu|(\Omega) := \sup_\pi \sum_{A \in \pi} \|\nu(A)\|_V \:,
\eeq
where the supremum is taken over all at most countable partitions~$\pi$ of~$\Omega$ into
disjoint $\nu$-measurable subsets. Clearly, the measure~$\nu$ is absolutely continuous
with respect to its variation measure. Therefore, we may form the Radon-Nikodym decomposition
\beq \label{RNdecomp}
d\nu(p) = A(p)\: d|\nu|(p) \qquad \text{with} \qquad A \in L^1 \big( \hat{K}, V; d|\nu| \big) \:.
\eeq
For almost all~$p \in \hat{K}$, the matrix~$A(p)$ is {\em{positive semi-definite}} on~$\la .|. \ra_V$ in the sense that
\[ \Sl u | A(p) u \Sr \geq 0 \qquad \text{for all~$u \in V$}\:. \]

Our goal is to form a spectral decomposition of the matrix~$A(p)$. In preparation, we 
recall a basic result on positive operators on indefinite
inner product spaces (more details can be found in the textbooks~\cite{bognar, GLR}).
It is useful to work in a pseudo-orthonormal basis.
In this basis, the indefinite inner product can be represented as
\beq \label{Sigdef}
\Sl .|. \Sr = \la\, . \,,\, S\, . \,\ra_{\C^{2n}} \:,
\eeq
where~$S$ is a diagonal matrix with entries~$\pm 1$. We also refer to~$S$ as
a {\em{signature operator}}.

\begin{Lemma} \label{lemmapositive}
Let~$A$ be a positive semi-definite linear operator on an indefinite inner product
space~$(V, \Sl .|. \Sr)$ of dimension~$2n$.
Then the spectrum of~$A$ is real.
Moreover, all the strictly positive (strictly negative) points in the spectrum correspond to
eigenspaces which are positive (respectively negative) definite.
\end{Lemma} 
\Proof In order to prove that the spectrum is real, we use the following perturbation argument.
For every~$\varepsilon>0$, the operator~$A+\varepsilon S$ is strictly positive.
Therefore, we can introduce a scalar product on~$V$ by
\[ \la .|. \ra_\varepsilon := \Sl \,. \,|\, (A+\varepsilon S)\, .\, \Sr \:. \]
Clearly, the operator~$A+\varepsilon S$ is symmetric with respect to this scalar product.
Using standard results from linear algebra, this operator is diagonalizable and has real eigenvalues.
Since the eigenvalues of a matrix depend continuously on the matrix entries, taking the limit~$\varepsilon
\searrow 0$, we conclude that also the operator~$A$ has a real spectrum.

Next, we use the spectral calculus to form an operator~$E_+$ which is symmetric, idempotent
and maps to the invariant subspaces corresponding to the strictly positive
eigenvalues. This operator can be defined for example as being the identity on all Jordan blocks
corresponding to the strictly positive spectral points and being zero otherwise.
Let~$V_+ := E_+(V)$ be the invariant subspace corresponding to all the strictly positive eigenvalues.
On this subspace, we introduce a positive semi-definite sesquilinear form~$\la .|. \ra_+$ by
\[ \la .|. \ra_+ := \Sl \,.\,|\, E_+ \,A \,.\, \Sr \:. \]
Let us show that this sesquilinear form
is even positive. To this end, let~$u$ be a non-zero vector in~$V_+$.
Since~$\Sl .|. \Sr$ is non-degenerate, there is a vector~$w \in V$ with~$\Sl u \,|\, w \Sr \neq 0$.
Using that~$A$ is invariant on~$V_+$ and invertible, there is~$v \in V_+$ with~$A v = E_+ w$. 
Then
\[ \la u \,|\, v \ra_+ = \Sl u \,|\, E_+ A v \Sr = \Sl u \,|\, E_+ w \Sr = \Sl u \,|\, w \Sr \neq 0 \:, \]
showing that~$\la .|. \ra_+$ is indeed a scalar product.

The operator~$E_+ A|_{V_+}$ is symmetric with respect to the scalar product~$\la .|. \ra_+$.
Hence it can be diagonalized.

Repeating this argument on the invariant subspace~$V_-$ corresponding to all the strictly negative eigenvalues,
we conclude that~$A$ is diagonalizable also on this invariant subspace.
It remains to show that these eigenspaces are definite. To this end, let~$u \in V_+ \cup V_-$
be an eigenvector, i.e.\ $A u = \lambda u$ and~$u \neq 0$. Then
\[ 0 < \Sl u | A u \Sr = \lambda\, \Sl u | u \Sr \:. \]
Therefore, the eigenspaces corresponding to the strictly positive 
(strictly negative) eigenvalues are indeed positive (respectively negative) definite.
\QED
We note for clarity that the point zero in the spectrum in general does {\em{not}} correspond to an eigenspace
of a positive semi-definite operator. A simple counter example is to choose~$n=1$ and
\[ S = \begin{pmatrix} 1 & 0 \\ 0 & -1 \end{pmatrix} \qquad \text{and} \qquad
A = \begin{pmatrix} 1 & 1 \\ -1 & -1 \end{pmatrix} \:. \]
The matrix~$A$ is positive semi-definite on~$V$, but it is nilpotent and not diagonalizable.

Applying Lemma~\ref{lemmapositive} to the operator~$A(p)$ in~\eqref{totvar}, we obtain a decomposition
\[ A(p) = A_+(p) + A_0(p) + A_-(p) \:, \]
where~$A_\pm$ are the restrictions to the strictly positive respectively negative spectral subspaces,
and~$A_0$ is the restriction to the invariant subspace corresponding to the spectral point zero.
The resulting decomposition of the measure~$\nu$ is denoted by
\beq \label{nudecomp}
\nu = \nu_+ + \nu_0 + \nu_- \:.
\eeq
Note that this decomposition is independent of the choice of the norm~$\|.\|_V$. Indeed,
choosing a different norm changes~$A(p)$ only by constant, which drops out when multiplying
by~$d|\nu|(p)$.

\begin{Def} \label{defnudecomp}
The measure~$\nu_+$ in~\eqref{nudecomp} is referred to as the {\bf{particle measure}}.
Likewise, $\nu_-$ is the {\bf{sea measure}}, and~$\nu_0$ is the {\bf{neutral measure}}.
\end{Def}

We finally explain these notions in the example of Dirac spinors mentioned in the introduction.
For the measure~\eqref{nuunreg}, the Radon-Nikodym decomposition gives (up to an irrelevant
constant depending on~$p$) the matrix
\[ A(p) = -(\slashed{p}+m) \:. \]
Clearly, this matrix is positive semi-definite. The computation
\[ \big( -(\slashed{p}+m) \big)^2 = p^2 + 2m \slashed{p} + m^2 = -2m\: \big( -(\slashed{p}+m) \big) \]
(where we used that~$\slashed{p}\slashed{p}= p^2 \1 =m^2 \1$)
shows that the eigenvalues of the matrix~$A(p)$ are negative. Moreover, in agreement with the general statement
of Lemma~\ref{lemmapositive}, its image is negative definite. Thus~\eqref{nuunreg} is
a sea measure. This is consistent with the fact that this measure describes the so-called Dirac sea
(for the physical concept of the Dirac sea see for example~\cite[Section~1.2]{cfs}
or~\cite[Sections~1.5 and~5.8]{intro}).
In order to describe particles, one needs to occupy states on the upper mass shell by choosing
\[ A(p) \sim \slashed{p}+m \qquad \text{with} \qquad p^2=m^2 \text{ and } p^0>0 \:, \]
giving rise to a particle measure.
The neutral measure can be thought of as a degenerate case where the sea
and particle subspaces have a non-trivial intersection. In a physical example, this happens
for massless Dirac particles, because in this case
\[ A(p) \sim \slashed{p} \qquad \text{and} \qquad p^2 = 0 \:. \]
Thus the matrix~$A(p)$ is nilpotent, giving rise to a neutral measure.

It is quite remarkable that already the abstract setup of the homogeneous causal action principle
incorporates a general notion of particles and anti-particles.

\section{Compactness and Existence of Minimizers} \label{secexist}
The existence of minimizers was first proven in~\cite[Section~4]{continuum}
in a slightly different setting (in particular, the dimension constraint was replaced
by a variant of what is now referred to as the boundedness constraint).
More recently, existence and compactness properties
were established in~\cite{langermom}. Here we give a short review,
keeping the presentation as simple as possible
by restricting attention to the methods and results needed later on.
We begin with a general compactness statement.

\begin{Thm} {\bf{(compactness)}} \label{thmcompact}
Let~$(\nu_k)_{k \in \N}$ be a sequence of positive definite measures
on the compact subset~$\hat{K} \subset \hscrM$ each of which satisfies the constraints~\eqref{TC} and~\eqref{DC}.
Then there is a sequence of unitary transformations~$(U_k)_{k \in \N}$ on the indefinite inner product
space~$(V, \Sl .|. \Sr)$ such that a subsequence~$\nu_{k_j}$ of the unitarily transformed measures converge weakly
(i.e.\ in the weak*-topology) to a positive definite measure~$\nu$, i.e.\
\[ U_{k_j} \,\nu_{k_j}\, U_{k_j}^* \rightarrow \nu \:. \]
The limit measure~$\nu$ again satisfies the constraints~\eqref{TC} and~\eqref{DC}.
Moreover, the action and the boundedness constraint are lower semi-continuous,
\[ \Sact(\nu) \leq \liminf_{k \rightarrow \infty} \Sact\big( \nu_k \big) \qquad \text{and} \qquad
 \T(\nu) \leq \liminf_{k \rightarrow \infty} \T\big( \nu_k \big) \:. \]
\end{Thm}
\Proof
The method of proof is similar to that of~\cite[Theorem~4.2(I)]{continuum}.
We let~$\|.\|_V$ be an arbitrarily chosen norm on~$V$.
Applying the perturbation argument in the proof of Lemma~\ref{lemmapositive},
the operator~$\nu(\hat{K}) + \varepsilon S$ is diagonalizable for any~$\varepsilon>0$.
Therefore~$\nu(\hat{K})$ is diagonalizable up to an arbitrarily small error term.
More precisely, we may choose a pseudo-orthonormal basis of~$(V, \Sl .|. \Sr)$ and
unitary operators~$U_k$ such that
\beq \label{Ukdecomp}
U_k\, \nu_k(\hat{K})\, U_k^* = \text{diag} \big( \alpha^{(k)}_1, \ldots, \alpha_{2n}^{(k)} \big) + \Delta B_n \:,
\eeq
where the~$\alpha_\ell^{(k)}$ are 
the eigenvalues of~$\nu_k(\hat{K})$ and
\[ \|\Delta B_k \|_V \leq \frac{1}{k} \qquad \text{for all~$k \in \N$} \]
(for more details see~\cite[Lemma~4.4]{continuum} or~\cite[proof of Lemma~4.2]{discrete}).
The dimension constraint~\eqref{DC} gives uniform bounds for the diagonal matrix in~\eqref{Ukdecomp}.
We conclude that the operators on the left side of~\eqref{Ukdecomp} are bounded uniformly in~$k$.
As a consequence, for every~$u \in V$, the positive definite measures~$d\Sl u | \nu_k u \Sr$ are uniformly bounded. Therefore, we can apply the
Banach-Alaoglu theorem and the Riesz representation theorem to conclude that
a subsequence converges in the weak*-topology to a positive definite measure~$d\Sl u | \nu u \Sr$.
Since the vector space~$V$ is finite-dimensional, choosing inductive subsequences, one can
arrange that the positive definite measures~$\nu_{k_\ell}$ converge to a positive definite measure~$\nu$.

The convergence as positive definite measures ensures that the constraints~\eqref{TC} and~\eqref{DC}
also hold for the limit measure~$\nu$.
The lower semi-continuity of the functionals~$\Sact$ and~$\T$ is an immediate consequence
of Fatou's lemma for sequences of measures.
\QED

Clearly, this compactness result also establishes the existence of minimizers.
We now state and prove the existence results needed here.
We again choose a pseudo-orthonormal basis~$(e_i)_{i=1,\ldots, 2n}$ of~$V$,
where the indefinite inner product takes the form~\eqref{Sigdef}.
\begin{Lemma} The dimension constraint is bounded from above by
\beq \label{sumlower}
\sum_{\ell=1}^{2n} \big|\alpha_\ell\big| \leq \Tr \big( S\, \nu(\hat{K}) \big) \:.
\eeq
\end{Lemma}
\Proof Similarly to~\eqref{Ukdecomp}, we can diagonalize~$\nu(\hat{K})$ up to an arbitrarily small
error term. More precisely, given~$\varepsilon>0$, we can diagonalize the
matrix~$\nu(\hat{K}) + \varepsilon S$, i.e.\
\beq \label{hdiag}
\nu(\hat{K}) + \varepsilon S = U D U^{-1} \:,
\eeq
where~$D \in \Lin(V)$ is diagonal and~$U$ is unitary on~$(V, \Sl .|. \Sr)$.
Since the eigenvalues depend continuously on~$\varepsilon$,
we conclude that for any~$\delta>0$ there is~$\varepsilon>0$ such that
\[ \sum_{\ell=1}^{2n} \big|\alpha_\ell\big| \leq \Tr(SD) + \delta \:. \]
Next, 
we use~\eqref{hdiag} to obtain
\[ \Tr \big( S\, \nu(\hat{K}) \big) = \Tr \big( S\, (\nu(\hat{K}) + \varepsilon S) \big)
- 2n\varepsilon = \Tr \big( U^{-1} S U\, D \big) - 2n\varepsilon \:. \]
The unitarity of~$U$ can be written as
\[ U^{-1} = S U^\dagger S \:, \]
giving the identity
\beq \label{id1}
\Tr \big( S\, \nu(\hat{K}) \big) = \Tr \big( U^\dagger U \:DS \big) - 2n\varepsilon \:.
\eeq
For every basis vector~$e_i$,
\[ \la e_i, U^\dagger U e_i \ra_{\C^{2n}} = \la U e_i, U e_i \ra_{\C^{2n}} \geq 
\big| \Sl U e_i \,|\, U e_i \Sr \big| = \big| \Sl e_i \,|\, e_i \Sr \big| = 1 \:. \]
Hence
\beq \label{in1}
\Tr \big( U^\dagger U \:DS \big) \geq  \Tr \big( DS \big)
= \Tr \big( SD \big) \geq \sum_{\ell=1}^{2n} \big|\alpha_\ell\big| - \delta \:.
\eeq
Combining~\eqref{id1} with~\eqref{in1}, we conclude that
\[ \Tr \big( S\, \nu(\hat{K}) \big) \geq \sum_{\ell=1}^{2n} \big|\alpha_\ell\big| - \delta - 2n\varepsilon\:. \]
Since~$\delta$ and~$\varepsilon$ can be chosen arbitrarily small, the result follows.
\QED

\begin{Thm} \label{thmexist}
Assume that~$0 < c < f$. Let~$(\nu_k)_{k \in \N}$ be a sequence of positive definite measures
on the compact subset~$\hat{K} \subset \hscrM$, which is a minimizing sequence of the
homogeneous causal action~\eqref{Sdef} and respects the trace constraint~\eqref{TC}
as well as the dimension constraint~\eqref{DC}.
Then there is a sequence of unitary transformations~$(U_k)_{k \in \N}$ on the indefinite inner product
space~$(V, \Sl .|. \Sr)$ such that a subsequence~$\nu_{k_j}$ of the unitarily transformed measures converge in the weak*-topology to a positive definite measure~$\nu$,
\beq \label{Ukchoice}
U_{k_j} \,\nu_{k_j}\, U_{k_j}^* \rightarrow \nu \:.
\eeq
The limit measure~$\nu$ is a minimizer of the causal action principle under the
trace constraint as well as the dimension constraint replaced by the
\begin{align}
\text{modified dimension constraint:} && \Tr \big( S\, \nu(\hat{K}) \big) &\leq f \:. \label{DCprime}
\end{align}
\end{Thm}
\Proof As in the proof of Theorem~\ref{thmcompact}, we diagonalize~$\nu_k(\hat{K})$ up
to an error of the order~$1/k$. Then the limit measure~$\nu$ has the property that~$\nu(\hat{K})$ is diagonal
and thus
\beq \label{nudiag}
\sum_{\ell=1}^{2n} \big|\alpha_\ell\big| = \Tr \big( S\, \nu(\hat{K}) \big) \:.
\eeq
In view of~\eqref{sumlower}, the inequality~\eqref{DCprime} is stronger than~\eqref{DC}.
Therefore, $\nu$ remains a minimizer if we replace the dimension constraint~\eqref{DC}
by the constraint~\eqref{DCprime}.
\QED

We conclude with two remarks. First, we point out that the signature operator~$S$
in the modified dimension constraint can be chosen arbitrarily, because the freedom in choosing~$S$
can be absorbed into the unitary operators~$U_k$ in~\eqref{Ukchoice}.
On the other hand, starting from a given minimizing measure~$\nu$, one obtains the
corresponding signature operator~$S$ satisfying~\eqref{DCprime} by choosing a pseudo-orthonormal
basis which diagonalizes~$\nu(\hat{K})$ (such a pseudo-orthonormal basis exists in view of~\eqref{nudiag}).
For this choice of~$S$, the modified dimension constraint~\eqref{DCprime} coincides with the 
dimension constraint~\eqref{DC}. We finally clarify the structure of the modified dimension constraint:

\begin{Remark} {\bf{(The modified dimension constraint)}} \label{remdim} {\em{
Using~\eqref{Pxidef} in~\eqref{DCprime}, the modified dimension constraint can be written as
\beq \label{DCP}
-\Tr \big( S\, P(0) \big) \leq f \:.
\eeq
For a Dirac sea configuration~\eqref{Punreg} and its spherically symmetric regularizations,
the operator~$S$ coincides with the Dirac matrix~$\gamma^0$. Then the left side of~\eqref{DCP}
becomes~$-\Tr(\gamma^0 P(x,x))$, having the interpretation as the total charge density at the
spacetime point~$x$ (including the states of the Dirac sea). For homogeneous systems, this
charge density coincides, up to a multiplicative constant, with the total charge.
More precisely, this multiplicative constant is the total spatial volume of the system.
It could be infinite, in which case one needs to take a suitable infinite volume limit.
For our purposes, it suffices to consider the total spatial volume as an irrelevant constant prefactor,
making it possible to identify the charge density with the total charge.
Using that every Dirac particle of a given Dirac sea has the same electric charge, the total charge coincides
(again up to an irrelevant constant) with the number of occupied states, which corresponds
to the dimension of the Hilbert space~$\H$. In this way, one gets a more direct connection
between the dimension constraint and the dimension of~$\H$.}} \QEDrem
\end{Remark}

\section{First Variations of the Homogeneous Causal Action} \label{secfirstvary}
We now consider first variations of the homogeneous causal action.
To this end, we consider a variation~$(\tilde{\nu}_\tau)_{\tau \in [0,\tau_{\max})}$ with~$\tau_{\max}>0$
of the measure~$\nu$ in the class of positive definite measures on~$\hat{K}$, i.e.\
\bitem
\item[{\rm{(i)}}] $\tilde{\nu}_0 = \nu$
\item[{\rm{(ii)}}] For every~$\tau \in [0,\tau_{\max})$, $\tilde{\nu}_\tau$ is a positive definite
measure on~$\hat{K}$ (see Definition~\ref{defposV}).
\eitem
First variations of the causal Lagrangian were computed in~\cite[\S1.4.1]{cfs}.
Under the assumption that the Lagrangian is differentiable in the direction of the variation, it
was shown that (see~[eq.~(1.4.16) and~eq.~(1.4.17)]\cite{cfs})
\beq \label{delLdef}
\delta \L(\xi) = 2\, \re
\Tr \big( Q(-\xi)\, \delta P(\xi) \big) \:,
\eeq
where~$\Tr$ denotes the trace on~$V$, and~$Q(x,y) : V \rightarrow V$ is a kernel
which is symmetric in the sense that
\beq \label{Qsymm}
Q(\xi)^* = Q(-\xi)\:.
\eeq
Integrating over~$\xi$ gives the variation of the homogeneous causal action~\eqref{Sdef}.
Before going on, we point out that the causal Lagrangian~\eqref{Ldef} in general is
{\em{not}} differentiable. Therefore, the formula~\eqref{delLdef} poses an implicit condition
on the admissible class of variations. Alternatively, one can proceed by approximating the causal Lagrangian~\eqref{Ldef} by smooth Lagrangians (for example obtained by mollification)
and take the limit of the resulting EL equations when the mollifier is removed
(see also the discussion in Section~\ref{secoutlook}).

For our purposes, it is most convenient to rewrite the first variation of the action in momentum space:
\begin{Prp} \label{prpvariation} Assume that the kernel~$Q(\xi)$ defined by~\eqref{delLdef}
is integrable,
\beq \label{QL1}
Q \in L^1(\scrM, d\nu) \:.
\eeq
 Then the first variation of the homogeneous causal action is given by
 \beq \label{delS}
 \delta \Sact := \frac{d}{d\tau} \Sact \big(\tilde{\nu}_\tau \big) \big|_{\tau=0}
= 2\:\frac{d}{d\tau}  \int_{\hat{K}} \Tr \Big( \hat{Q}(p)\: d\tilde{\nu}_\tau(p) \Big) \bigg|_{\tau=0}
\eeq
with
\[ \hat{Q}(p) := \int_{\scrM} Q(\xi)\: e^{-i p \xi}\: d\mu(\xi) \;\in\; C^0(\hscrM, \Lin(V)) \:. \]
\end{Prp}
\Proof Using~\eqref{Pxidef} in~\eqref{delLdef}, we obtain
\[ \delta \L(\xi) = 2\, \re  \int_{\hat{K}} e^{-i p \xi} \:
\Tr \big( Q(-\xi)\, d\delta \nu(p) \big) \:. \]
Integrating over~$\xi$ gives
\[ \delta \Sact = 2\, \re  \int_{\scrM}  d\mu(\xi) \int_{\hat{K}} e^{-i p \xi} \:
\Tr \big( Q(-\xi)\, d\delta \nu(p) \big)\:. \]
Using that~$Q$ is integrable, we may interchange the integrals to obtain
\beq \label{dSreal}
\delta \Sact = 2\, \re  \int_{\hat{K}} \Tr \big( \hat{Q}(p)\, d\delta \nu(p) \big)\:.
\eeq
The symmetry of the kernel~$Q$, \eqref{Qsymm}, implies that its Fourier transform is symmetric
(with respect to the inner product~$\Sl .|. \Sr$), i.e.\
\beq \label{Qpsymm}
\hat{Q}(p)^* = \hat{Q}(p)\:.
\eeq
Therefore, we may leave out the real part in~\eqref{dSreal}, giving~\eqref{delS}.

It remains to show that~$\hat{Q}(p)$ is continuous. To this end, we estimate the
difference of Fourier integrals by
\begin{align*}
\big\| \hat{Q}(p') - \hat{Q}(p) \big\|_{\Lin(V)} \leq 
\int_{\scrM} \| Q(\xi) \|_{\Lin(V)}\: \big| e^{-i p' \xi} - e^{-i p \xi} \big|\: d\mu(\xi)
\end{align*}
Clearly, in the limit~$p' \rightarrow p$ the integrand converges to zero pointwise. Moreover,
the integrand is dominated by the integrable function~$2 \,\| Q(\xi) \|_{\Lin(V)} \in L^1(\scrM)$.
Therefore, Lebesgue's dominated convergence yields
\[ \lim_{p' \rightarrow p} \big\| \hat{Q}(p) - \hat{Q}(p') \big\|_{\Lin(V)} = 0 \:. \]
This concludes the proof.
\QED

\section{Reformulation in Terms of Measures on~$\Q \subset \Symm(V)$} \label{secpush}
The remaining difficulty is that the first variations as computed in Proposition~\ref{prpvariation}
must satisfy all the constraints. Due to this difficulty, it is not obvious what the variational formula in Proposition~\ref{prpvariation}
tells us about~$\hat{Q}(p)$.
Before entering the detailed analysis of this problem, in this section we simplify the variational formula~\eqref{delS}.
The idea is to introduce~$q:=\hat{Q}(p)$ as a new integration variable. This has the advantage that
the integrand becomes particularly simple. This reformulation also has the major advantage
that it makes it possible to generalize our methods to situations in which~$\hat{Q}(p)$ is no longer continuous.

We denote the symmetric linear operators on~$V$ by~$\Symm(V) \subset \Lin(V)$,
where ``symmetric'' refers to the inner product~$\Sl .|. \Sr$, i.e.\ to the condition
\[ \Sl A u | v\Sr = \Sl u | A\, v \Sr \qquad \text{for all~$u, v \in V$}\:. \]
In the proof of Proposition~\ref{prpvariation} we saw that~$\hat{Q}(p)$ is symmetric, \eqref{Qpsymm}.
We thus obtain the continuous mapping
\[ \hat{Q} \in C^0\big( \hat{K}, \Symm(V) \big) \:. \]
Clearly, the image of this mapping is again compact; we denote it by
\[ \Q := \hat{Q} \big( \hat{K} \big) \subset \Symm(V) \qquad \text{compact} \:. \]
We let~$\mu$ be the push-forward of~$\nu$ under the mapping~$\hat{Q}$. It is a positive
definite measure supported on~$\Q$,
\[ \mu := \hat{Q}_* \nu \:,\qquad \supp \mu \subset \Q \:. \]
Moreover, by definition of the push-forward measure, 
\[ \int_{\hat{K}} \Tr \Big( \hat{Q}(p)\: d\nu(p) \Big)
= \int_{\Q} \Tr \Big( q\: d\mu(q) \Big) \:. \]
Therefore, Proposition~\ref{prpvariation} can be rephrased in terms of~$\mu$ as follows.
\begin{Prp} Assume that the kernel~$Q(\xi)$ is integrable, \eqref{QL1}.
Then the first variation of the homogeneous causal action is given by
\[ 
 \delta \Sact = 
 2\:\frac{d}{d\tau}  \int_{\Q} \Tr \Big( q\: d\tilde{\mu}_\tau(q) \Big) \bigg|_{\tau=0} \:. \]
\end{Prp} \noindent
Moreover, again by definition of the push-forward measure, we know that
\[ \mu(\Q) = \nu(\hat{K}) \:, \]
making it possible to formulate also the trace constraint~\eqref{TC} and
the modified dimension constraint~\eqref{DCprime}
in terms of~$\mu$ by
\beq \label{TDC}
\Tr \big( \mu(\Q) \big) = c \qquad \text{and} \qquad
\Tr \big( S\, \mu(\Q) \big) \leq f \:.
\eeq

\section{The Euler-Lagrange Equations} \label{secEL}

\subsection{Treating the Scalar Constraints} \label{secscalconstr}
The main difficulty in the derivation of the Euler-Lagrange equations are the constraints.
Apart from the trace and modified dimension constraints~\eqref{TC} and~\eqref{TDC},
the measure~$\nu$ must be positive definite (see Definition~\ref{defposV}),
which can be understood as an infinite number
of inequality constraints. In view of these inequality constraints, the Lagrange multiplier method
cannot be used directly. Instead, our strategy is to satisfy the constraints~\eqref{TDC} explicitly
by variations within the class of positive definite measures.
For ease in notation, we refer to the trace constraint and the modified dimension constraint
as the {\em{scalar constraints}}. We again work in the basis where~$S$ is diagonal and use the block matrix notation
\[ S = \begin{pmatrix} \1 & 0 \\ 0 & -\1 \end{pmatrix} \:. \]
For~$s_1,s_2 \in (-1,1)$ we consider the family of measures
\beq \label{s12vary}
\mu_{s_1,s_2} := \begin{pmatrix} (1+s_1) \1 & 0 \\ 0 & (1+s_2) \1 \end{pmatrix}
\mu \begin{pmatrix} (1+s_1) \1 & 0 \\ 0 & (1+s_2) \1 \end{pmatrix} \:.
\eeq
Since we multiply from the left and right by a symmetric operator on~$(V, \Sl .|. \Sr)$, this family of measures
is again positive definite. The first variation of the scalar constraints is computed by
\begin{align}
\frac{\partial}{\partial s_1} \Tr \big( S\, \mu_{s_1,s_2}(\Q) \big) \big|_{s_1=s_2=0}
= \frac{\partial}{\partial s_1} \Tr \big( \mu_{s_1,s_2}(\Q) \big) \big|_{s_1=s_2=0}
&= \Tr \big( S\, \mu(\Q) \big) + c \label{par1} \\
\frac{\partial}{\partial s_2} \Tr \big( S\, \mu_{s_1,s_2}(\Q) \big) \big|_{s_1=s_2=0}
=-\frac{\partial}{\partial s_2} \Tr \big( \mu_{s_1,s_2}(\Q) \big) \big|_{s_1=s_2=0}
&= \Tr \big( S\, \mu(\Q) \big) - c \:. \label{par2}
\end{align}

We now distinguish the following cases: 
\label{cases}
\bitem
\item[{\bf{(a)}}] $c \leq \Tr \big( S\, \mu(\Q) \big)<f$: \\[0.2em]
In this case, the dimension constraint can be disregarded. In order to satisfy the trace constraint,
it suffices to consider rescalings, i.e.\ variations of the form~\eqref{s12vary} with~$s_1=s_2$.
\item[{\bf{(b)}}] $\Tr \big( S\, \mu(\Q) \big) = f$: \\[0.2em]
Since we assume~$c<f$ throughout this paper, it follows that
\[ c = \Tr \big( \mu(\Q) \big) \;<\; \Tr \big( S\, \mu(\Q) \big) = f \:. \]
Therefore, we can preserve both equations~$\Tr \big( \mu(\Q) \big) = c$
and~$\Tr \big( S\, \mu(\Q) \big) = f$ in the variation by choosing~$s_1$ and~$s_2$ appropriately.
\eitem

\subsection{Variations Changing the Support} \label{secchange}
An operator~$A \in \Symm(V)$ is said to be {\em{positive semi-definite}} if
\[ \Sl u | A u \Sr \geq 0 \qquad \text{for all~$u \in V$}\:. \]
We denote the set of all positive semi-definite operators by~$\Symm_+(V)$.
We let~$q \in \Q$ and choose an arbitrary positive semi-definite operator~$A \in \Symm_+(V)$.
We consider the variation~$(\tilde{\mu}_\tau)_{\tau \in [0,1)}$ with
\[ \tilde{\mu}_{\tau} = \mu_{s_1,s_2} + \tau\, A \: \delta_q \]
with~$\mu_{s_1, s_2}$ as in~\eqref{s12vary}, 
where~$s_1$ and~$s_2$ are linear functions in~$\tau$,
\beq \label{s12choice}
s_1 = \kappa_1 \,\tau \:,\qquad s_1 = \kappa_2 \,\tau \:.
\eeq
Note that~$q$ does not need to be in the support of~$\mu$, in which case the
support of the measure changes in the variation.
Using~\eqref{par1} and~\eqref{par2}, the first variations of the constraints are computed by
\begin{align*}
\frac{d}{d\tau} \Tr \big(\tilde{\mu}_{\tau}(\Q) \big) \big|_{\tau=0} 
&= \Big( \kappa_1 \frac{\partial}{\partial s_1} + \kappa_2 \frac{\partial}{\partial s_2} \Big)
\Tr \big(\mu_{s_1,s_2}(\Q) \big) \big|_{s_1=s_2=0} + \Tr (A) \\
&= (\kappa_1 + \kappa_2)\: c + (\kappa_1-\kappa_2)\: \Tr \big( S\,\mu_{s_1,s_2}(\Q) \big) + \Tr (A) \\
\frac{d}{d\tau} \Tr \big(S\, \tilde{\mu}_{\tau}(\Q) \big) \big|_{\tau=0}
&= \Big( \kappa_1 \frac{\partial}{\partial s_1} + \kappa_2 \frac{\partial}{\partial s_2} \Big) 
\Tr \big(S\,\mu_{s_1,s_2}(\Q) \big) \big|_{s_1=s_2=0} + \Tr (SA) \\
&= (\kappa_1 + \kappa_2)\: \Tr \big(S\, \mu_{s_1,s_2}(\Q) \big) + (\kappa_1-\kappa_2)\: c + \Tr (S A) \:.
\end{align*}
Our strategy is to choose the parameters~$\kappa_1$ and~$\kappa_2$ such that the
constraints are preserved in first variations. 
More precisely, in case~{\bf{(a)}} on page~\pageref{cases}, we only need to satisfy the
trace constraint. This can be arranged by choosing
\beq \label{kappa12a}
\kappa_1 = \kappa_2 = -\frac{1}{2c} \:\Tr(A) \:.
\eeq
In case~{\bf{(b)}} on page~\pageref{cases}, on the other hand, we need to
arrange that both scalar constraints are preserved in first variations. To this end, we choose
\beq \label{kappa12b}
\kappa_1 = -\frac{\Tr(SA)+\Tr(A)}{f+c}\:,\qquad \kappa_2 = -\frac{\Tr(SA)+\Tr(A)}{f-c} \:.
\eeq
We remark that, having satisfied the constraints for first variations, they can also be
satisfied nonlinearly for small~$\tau$ by employing the implicit function theorem.

Having satisfied the scalar constraints,
the homogeneous action is minimal under first variations of~$\tilde{\mu}_\tau$.
A short computation yields
\begin{align*}
0 &\geq \frac{d}{d\tau}  \int_{\Q} \Tr \Big( \tilde{q}\: d\tilde{\mu}_\tau(\tilde{q}) \Big) \bigg|_{\tau=0} \\
&= \Big( \kappa_1 \frac{\partial}{\partial s_1} + \kappa_2 \frac{\partial}{\partial s_2} \Big)
\int_{\Q} \Tr \Big( \tilde{q}\: d\mu_{s_1, s_2}(\tilde{q}) \Big)  + \Tr (q A) \\
&= (\kappa_1+\kappa_2) \int_{\Q} \Tr \Big( \tilde{q}\: d\mu(\tilde{q}) \Big)
+ (\kappa_1-\kappa_2) \int_{\Q}\: \Tr \Big( \frac{1}{2}\:\{\tilde{q}, S\}\: d\mu(\tilde{q}) \Big) + \Tr (q A) \:.
\end{align*}
Using the above form of the parameters~$\kappa_1$ and~$\kappa_2$, we obtain the following result:
\begin{Lemma} \label{lemmachange}
Let~$\nu$ be a positive definite measure on~$\hat{K}$ which is
minimizer of the homogeneous causal action principle under the trace constraint~\eqref{TC}
and the modified dimension constraint~\eqref{DCprime}.
Moreover, let~$\mu = \hat{Q}_* \nu$ be the corresponding measure on~$\Q \subset \Symm(V)$.
Then, for a suitable choice of Lagrange parameters~$\alpha, \beta \in \R$,
the following variational inequality holds,
\[ \Tr \big( (q - \alpha\, \1 - \beta\, S)\: A\big) \geq 0 \qquad \text{for all~$q \in \Q$ and~$A \in \Symm_+(V)$} \:. \]
More precisely, in case~{\rm{\bf{(a)}}} on page~\pageref{cases}, 
the Lagrange parameters are given by
\beq \label{alphabetaa}
\alpha = \frac{1}{c} \: \int_{\Q} \Tr \Big( \tilde{q}\: d\mu(\tilde{q}) \Big) \:,\qquad
\beta = 0 \:,
\eeq
whereas in case~{\rm{\bf{(b)}}} on page~\pageref{cases} they are given by
\begin{align}
\alpha &= \frac{1}{f^2-c^2} \: \bigg( f \int_{\Q}\: \Tr \Big( \frac{1}{2}\:\{\tilde{q}, S\}\: d\mu(\tilde{q}) \Big) 
- c \int_{\Q} \Tr \Big( \tilde{q}\: d\mu(\tilde{q}) \Big) \bigg)\:, \label{alphadef} \\
\beta &= \frac{1}{f^2-c^2} \: \bigg( f \int_{\Q} \Tr \Big( \tilde{q}\: d\mu(\tilde{q}) \Big)
- c \int_{\Q}\: \Tr \Big( \frac{1}{2}\:\{\tilde{q}, S\}\: d\mu(\tilde{q}) \Big) \bigg) \:. \label{betadef}
\end{align}
\end{Lemma}

\subsection{Variations With Fixed Support} \label{secfix}
Let~$q \in \Q$ and~$U \subset \Symm(V)$ an open neighborhood of~$q$. We consider
the variation~$(\tilde{\mu}_\tau)_{\tau \in (-1,1)}$ with
\[ \tilde{\mu}_\tau = \mu_{s_1, s_2} + \tau\, \chi_U \mu \:. \]
In order to satisfy the scalar constraints, we choose~$s_1$ and~$s_2$ again as in~\eqref{s12choice}
with~$\kappa_1$ and~$\kappa_2$ again according to~\eqref{kappa12a} or~\eqref{kappa12b},
however with the obvious replacement
\[ A \rightarrow \mu(U) \:. \]
We point out that, in contrast to the variations in the previous section, now we may choose~$\tau$ negative,
giving rise to a variational {\em{equality}}. Similar as explained at the beginning of Section~\ref{secdecomp},
we form the Radon-Nikodym decomposition
\beq \label{radon2}
d\mu(q) = A(q)\: d|\mu|(q) \qquad \text{with} \qquad
A(q) \in \Symm_+(V) \:,
\eeq
where the absolute value again denotes the variation measure~\eqref{totvar}.

\begin{Lemma} \label{lemmafix}
Under the assumptions of Lemma~\ref{lemmachange} and for the same values of the
Lagrange parameters~$\alpha, \beta \in \R$,
the matrix~$A(q)$ in the Radon-Nikodym representation~\eqref{radon2} has the property
\[ \Tr \big( (q - \alpha\,\1 - \beta\, S)\: A(q) \big) = 0 \qquad \text{for almost all~$q \in \supp \mu$} \:. \]
\end{Lemma}
\Proof We have
\begin{align*}
0 &= \frac{d}{d\tau} \bigg( \int_{\Q} \Tr \Big( q\: d\tilde{\mu}_\tau(q) \Big)
- \alpha \,\Tr \big( \tilde{\mu}_\tau(\Q) \big) - \beta \: \Tr \big( S \tilde{\mu}_\tau(\Q) \big) \bigg) \bigg|_{\tau=0} \\
&= \int_U \Tr \Big( q\: d\mu(q) \Big) + \Tr \big( -\alpha \mu(U) - \beta\, S \mu(U) \big) \:.
\end{align*}
Using the Radon-Nikodym decomposition~\eqref{radon2}, we obtain
\[ 0 = \int_U \Tr \big( (q - \alpha\,\1 - \beta\, S)\: A(q) \big)\: d|\mu|(q) \:. \]
Since~$U$ is arbitrary, the result follows.
\QED

\subsection{Statement of the Euler-Lagrange Equations}
The Euler-Lagrange equations are obtained by combining the inequality in Lemma~\ref{lemmachange}
with the equality in Lemma~\ref{lemmafix}. The only shortcoming
is that these lemmas make a statement only on the trace of the 
operator product~$(q - \alpha \1 - \beta S)\, A(q)$. In order to get corresponding statements
for the operator product itself, we can use the following variant of the Cauchy-Schwarz inequality.

\begin{Lemma} \label{lemmacsu}
For every~$B \in \Symm(V)$, the following statement holds:
\[ \Tr(A B) \geq 0 \quad \text{for all~$A \in \Symm_+(V)$} \qquad \Longleftrightarrow \qquad
B \in \Symm_+(V) \:. \]
Moreover, for any~$A, B \in \Symm_+(V)$,
\[ \Tr(AB) = 0 \qquad \Longrightarrow \qquad A B = 0 \:. \]
\end{Lemma}
\Proof In order to relate the statements to elementary results in linear algebra,
it is convenient to again work in a pseudo-orthonormal basis where the indefinite inner product
is represented according to~\eqref{Sigdef}. Then~$A$ is symmetric on~$V$ if and only if~$SA$
is a Hermitian matrix. Likewise, $A$ is positive on~$V$ if and only if~$SA$ is a positive semi-definite matrix.
Similarly~$B$ is symmetric and positive on~$V$ if and only if~$BS$ is a Hermitian and positive semi-definite
matrix, respectively. Using that~$\Tr(AB) = \Tr((SA) (BS))$, the claim follows immediately from
corresponding statements for symmetric and positive matrices.
\QED

Combining this lemma with the statements of Lemmas~\ref{lemmachange} and~\ref{lemmafix}
gives the following result.
\begin{Lemma} \label{lemmaEL}
Let~$\nu$ be a positive definite measure on~$\hat{K}$ which is a
minimizer of the homogeneous causal action principle under the trace constraint~\eqref{TC}
and the modified dimension constraint~\eqref{DCprime}.
Moreover, let~$\mu = \hat{Q}_* \nu$ be the corresponding measure on~$\Q \subset \Symm(V)$.
Then, for the Lagrange parameters~$\alpha$ and~$\beta$ given by~\eqref{alphabetaa}
or~\eqref{alphadef} and~\eqref{betadef} the following statements hold:
\bitem
\item[{\rm{(i)}}] For every~$q \in \Q$, the operator~$q - \alpha \1 - \beta S$ is positive semi-definite
on~$V$.
\item[{\rm{(ii)}}] The matrix~$A(q)$ in the Radon-Nikodym representation~\eqref{radon2} has the property
\[ (q - \alpha\,\1 - \beta\, S)\: A(q) = 0 \qquad \text{for almost all~$q \in \supp \mu$} \:. \]
\eitem
\end{Lemma}

Our main theorem is obtained by rewriting the last result in terms of the
minimizing measure~$\nu$.
\begin{Thm} \label{thmEL}
Let~$\nu$ be a positive definite measure on~$\hat{K}$ which is
minimizer of the homogeneous causal action principle under the trace constraint~\eqref{TC}
and the modified dimension constraint~\eqref{DCprime}.
Moreover, let~$\mu = \hat{Q}_* \nu$ be the corresponding measure on~$\Q \subset \Symm(V)$.
Then, for the Lagrange parameters~$\alpha$ and~$\beta$ given by~\eqref{alphabetaa}
or~\eqref{alphadef} and~\eqref{betadef} the following statements hold:
\bitem
\item[{\rm{(i)}}] For every~$p \in \hat{K}$, the operator~$\hat{Q}(p) - \alpha \1 - \beta S$ is positive semi-definite
on~$V$.
\item[{\rm{(ii)}}] The following measures vanish,
\[ \big( \hat{Q}(p) -\alpha\,\1 - \beta\, S \big) \:d\nu(p) = 0 = 
d\nu(p) \:\big( \hat{Q}(p) -\alpha\,\1 - \beta\, S \big) \:. \]
\eitem
\end{Thm}

We finally formulate the EL equations in terms of a minimality property on the support.
\begin{Corollary} \label{thmELalt}
For any~$p \in \hat{K}$, we define
\[ g(p) := \inf \big\{ \lambda \geq 0 \:\big|\: \hat{Q}(p) - \alpha\,\1 - \beta\, S - \nu \text{ is positive semi-definite
for all~$\nu$ with~$|\nu|\leq \lambda$} \big\} \:. \]
Then
\[ g|_{\supp \nu} \equiv \inf_{p \in \hat{K}} g(p) = 0 \:. \]
\end{Corollary} \noindent
Applying Lemma~\ref{lemmapositive}, the function~$g(p)$ can also be understood as the maximal
size of the interval~$[-\lambda, \lambda]$ with the properties that
\[ (-\lambda, \lambda) \cap \sigma\big( \hat{Q}(p) - \alpha\,\1 - \beta\, S \big) = \varnothing \:, \]
and that all spectral points strictly larger than~$\lambda$ (strictly smaller than~$-\lambda$) correspond to
positive definite (respectively negative definite) eigenspaces.

\subsection{An Inequality for the Lagrange Parameters}
In our method for treating the scalar constraints introduced in Section~\ref{secscalconstr}
we arranged that the functionals in these constraints were constant.
In the case~$\Tr ( S\, \mu(\Q) )<f$, we only took into account the trace constraint,
whereas in the case~$\Tr ( S\, \mu(\Q) )=f$ we arranged that both~$\Tr ( \mu(\Q) )$
and~$\Tr ( S\, \mu(\Q) )$ were kept fixed.
The latter procedure does not take into account that the dimension constraint is
an {\em{inequality}} constraint. Thus, in the case~$\Tr ( S\, \mu(\Q) )=f$ it is possible
to consider variations which {\em{decrease}}~$\Tr(S\, \mu(\Q))$.
Such variations reveal that the Lagrange parameter~$\beta$ is always non-positive:
\begin{Prp}
Under the assumptions of Theorem~\ref{thmEL}, the Lagrange multiplier~$\beta$ is
always non-negative,
\beq \label{betaneg}
\beta \leq 0 \:.
\eeq
\end{Prp}
\Proof In case~{\rm{\bf{(a)}}} on page~\pageref{cases}, the inequality~\eqref{betaneg}
holds simply because the parameter~$\beta$ vanishes.
Therefore, it remains to consider case~{\rm{\bf{(b)}}} on page~\pageref{cases}, i.e.\
\beq \label{trseq}
\Tr \big( S\, \mu(\Q) \big) = f \:,
\eeq
We consider the variation
\[ \tilde{\mu}_\tau = \mu_{s_1, s_2} \]
with~$\mu_{s_1, s_2}$ as in~\eqref{s12vary} and the parameters~$s_1$ and~$s_2$ given by
\[ 
s_1 = - \tau \big( \Tr \big( S\, \mu(\Q) \big) - c \big) \:, \quad s_2 =  - \tau \big( \Tr \big( S\, \mu(\Q) \big) + c \big) \:. \]
This variation is admissible for sufficiently small~$\tau \geq 0$, because the trace constraint is respected,
whereas the dimension constraint is {\em{de}}creased,
\begin{align*}
\frac{d}{d\tau} \Tr \big( \tilde{\mu}_\tau(\Q) \big) \big|_{\tau=0} &= 0 \\
\frac{d}{d\tau} \Tr \big( S\,\tilde{\mu}_\tau(\Q) \big) \big|_{\tau=0} 
&=-2 \,\big( \Tr \big( S\, \mu(\Q) \big)^2 - c^2 \big) =-2 \,\big( f^2 - c^2 \big) < 0
\end{align*}
(in the last line we used~\eqref{trseq} and~\eqref{cfrange}).
Moreover,
\begin{align*}
0 &\leq \frac{d}{d \tau} \int_{\Q} \Tr \Big( q\: d\tilde{\mu}_\tau(q) \Big)\Big|_{\tau=0} \\
&= -2 f \int_{\Q} \Tr \Big( q\: d\mu(q) \Big)
+ c\: \int_{\Q} \Tr \Big( \{q\, S\}\, d\mu(q) \Big) \:.
\end{align*}
Using that
\[ \big( q -\alpha\,\1 - \beta\, S \big) \:d\mu(q) = 0 \:, \]
we conclude that
\begin{align*}
0 &\leq -2f \int_{\Q} \Tr \Big( (\alpha\,\1+\beta\, S)\: d\mu(q) \Big)
+ 2c\: \int_{\Q} \Tr \Big( (\alpha S + \beta) \, d\mu(q) \Big) \\
&= -2f \, \big( \alpha c + \beta f \big) 
+ 2c\: \big( \alpha f + \beta c \big) = - 2\beta\: \big(f^2-c^2 \big) \:.
\end{align*}
Since~$(f^2-c^2)>0$, it follows that~$\beta \leq 0$.
\QED

\section{A Method for Computing the Lagrange Parameters} \label{secpointwise}
In the previous computations, the Lagrange parameters~$\alpha$ and~$\beta$ were
determined by arranging the scalar constraints by a suitable variation of the measure
(see the variation~\eqref{s12vary} as well as~\eqref{alphabetaa} and~\eqref{alphadef}, \eqref{betadef}).

We now explain that the Lagrange parameters are even determined ``pointwise'' in the following sense:
\begin{Prp} \label{prp91} Let~$\nu$ be a minimizing measure and~$p \in \supp \nu$.
We let~$A=A(p)$ be the operator in the Radon-Nikodym decomposition~\eqref{RNdecomp}.
Assume that the strict inequality
\beq \label{strictpoint}
\big| \Tr \big( A(p) \big) \big| < \Tr \big( S A(p) \big)
\eeq
holds. Then the Lagrange parameters~$\alpha$ and~$\beta$
in the statement of Theorem~\ref{thmEL} are uniquely determined by demanding
that the conditions~{\rm{(i)}} and~{\rm{(ii)}} in this theorem hold at~$p$.
\end{Prp}

The proof of this proposition will be based on a variational principle which will
be introduced and studied in the next section. The proof will be
completed at the end of Section~\ref{secvarypoint}.

\subsection{A Pointwise Variational Principle} \label{secvarypoint}
For the proof of Proposition~\ref{prp91},
we shall set up and solve a variational principle
defined pointwise at~$q:=\hat{Q}(p)$. This variational principle will also shed some light on
the possible form of the operator~$A(p)$.
More generally, given~$q \in \Symm(V)$ and parameters~$a,b \in \R$,
we consider the following {\em{pointwise variational principle}}:
It is convenient to work in a pseudo-orthonormal basis~$(e_i)_{i=1,\ldots, 2n}$ of~$V$.
In this basis, the indefinite inner product can be represented again in the form~\eqref{Sigdef},
where~$S$ is a diagonal matrix with entries~$\pm 1$. We now
\beq \label{auxmin}
\text{minimize} \qquad \Tr (q A)
\eeq
under variations of~$A \in \Symm_+(V)$ in the class of all positive semi-definite operators, subject to the constraints
\beq \label{auxconstr}
\Tr(A) = a \qquad \text{and} \qquad \Tr(S A) = b \:.
\eeq

\begin{Lemma} \label{lemmaL1}
Given~$q \in \Symm(V)$, assume that~$a$ and~$b$ are in the range
\beq \label{abrange}
|a| \leq b \:.
\eeq
Then the variational problem~\eqref{auxmin} under the constraints~\eqref{auxconstr}
has a minimizer. This minimizer has the property that for suitable~$\alpha, \beta \in \R$,
\beq \label{Arel}
A\,\big(q  - \alpha\, \1 - \beta\, S \big) = 0 \:.
\eeq
\end{Lemma}
\Proof In order to parametrize the positive semi-definite operators on~$V$, it is most convenient to 
write~$A$ as
\beq \label{Arep}
A = S\, M^\dagger M \:,
\eeq
where~$M \in \Lin(\C^{2n})$ is any $2n\times 2n$-matrix, and the dagger denotes its Hermitian conjugate
(here we make use of the fact that~$SA$ is a positive semi-definite matrix, making it possible to
take its square root).
Then the auxiliary variational principle becomes
\beq \label{auxMM}
\text{minimize} \qquad \Tr \big(q \,S\, M^\dagger M \big) \:,
\eeq
where we vary~$M \in \Lin(\C^{2n})$ under the only constraints
\beq \label{newconstr}
\Tr \big( S\, M^\dagger M \big) = a \qquad \text{and} \qquad \Tr \big(M^\dagger M \big) = b \:.
\eeq
Moreover, applying the Cauchy-Schwarz inequality
\beq \label{cs}
\big| \Tr \big( S\, M^\dagger M \big) \big| \leq \|S\|\: \Tr \big( M^\dagger M \big) \leq b \:,
\eeq
the left equation in~\eqref{newconstr} can be satisfied if and only if~$|a| \leq b$. We thus obtain the
admissible parameter range~\eqref{abrange}.
For any admissible parameter values, the right equation in~\eqref{newconstr} shows that~$M$
may be varied only inside a compact set. Therefore, minima exist by continuity.

Let~$M$ be a minimizer. The corresponding EL equations can be obtained with the help of the
Lagrange multiplier rule. We first apply it naively and justify it afterward.
Adding multiples of the constraints to the first variation gives
\begin{align*}
0 &= \delta \Tr \big(q \,S\, M^\dagger M \big) - \alpha\: \delta \Tr \big( S\, M^\dagger M \big)
- \beta\: \delta \Tr \big(M^\dagger M \big) \\
&= 2 \re \Big( \Tr \big((\delta M)^\dagger M\,q \,S \big) - \alpha\: \delta \Tr \big( (\delta M^\dagger) M\,S \big)
- \beta\: \delta \Tr \big((\delta M^\dagger) M \big) \Big)
\end{align*}
(here we used that the matrix~$qS$ is Hermitian, because~$q \in \Symm(V)$
and~$(qS)^\dagger = S q^\dagger = S S q^* S = qS$).
Since~$\delta M$ can be chosen as an arbitrary $(2n \times 2n)$-matrix,
it follows that
\beq \label{mqS}
M\,q \,S  - \alpha\, M S - \beta\, M = 0 \:.
\eeq
Multiplying from the left by~$S M^\dagger$ and using~\eqref{Arep} gives
\[ A\,q \,S  - \alpha\, A S - \beta\, A = 0 \:. \]
Finally, we multiply from the right by~$S$ to obtain~\eqref{Arel}.

It remains to justify the Lagrange multiplier method. Computing the first variations of the constraints
\beq \label{fvc}
\begin{split}
\delta \Tr \big( M^\dagger M \big) &= 2 \re \Tr \big( (\delta M)^\dagger M \big) \\
\delta \Tr \big( S\, M^\dagger M \big) &= 2 \re \Tr \big( (\delta M)^\dagger M S \big) \:,
\end{split}
\eeq
one sees that the constraints are regular unless~$M$ is a multiple of~$MS$.
Multiplying the equation~$M = \kappa MS$ by~$S$, one sees that~$\kappa = \pm 1$, so that
\beq \label{Mrel}
M = \pm MS \:.
\eeq
Using this equation in~\eqref{newconstr}, it follows that~$b=\pm a$. If~$b=a=0$, then~$M=0$
is the trivial minimizer. In this case, also~\eqref{Arel} holds trivially with~$A=0$.
In the remaining case~$b=\pm a > 0$, the inequality in~\eqref{cs} becomes an equality. This means that
all matrices satisfying the constraints~\eqref{auxconstr} also satisfy the relation~\eqref{Mrel}.
Therefore, the two constraints in~\eqref{auxconstr} are multiples of each other. Thus we may
drop the first constraint. The remaining second constraint is regular, making it possible to apply the
Lagrange multiplier method. This gives~\eqref{Arel} with~$\alpha=0$.
\QED

\begin{Lemma} \label{lemmaminpos}
The operator~$q  - \alpha \1 - \beta\, S$ in~\eqref{Arel} is positive semi-definite.
\end{Lemma}
\Proof We again work in the pseudo-orthonormal basis of~$V$ where the indefinite inner product
has the representation~\eqref{Sigdef}. Then the claim is equivalent to the statement that the matrix
\[ N := q\,S  - \alpha\, S - \beta\, \1 \]
is positive semi-definite on~$(\C^{2n}, \la .,. \ra_{\C^{2n}})$.

This statement is proved as follows. Noting that the variational principle considered in
the proof of Lemma~\ref{lemmaL1} involves only~$M^\dagger M$
(see~\eqref{auxMM} and~\eqref{newconstr}), we can choose~$M=\sqrt{M^\dagger M}$.
For notational convenience, we unitarily transform~$\C^{2n}$ such that this matrix is diagonal.
Using a block matrix notation in the image of~$M$ and its orthogonal complement, we obtain
\[ M = \begin{pmatrix} X & 0 \\ 0 & 0 \end{pmatrix} \:, \]
where the matrix~$X$ is positive definite. With this notation, the identity~\eqref{mqS} means that~$N$
has the form
\beq \label{Nform}
N = \begin{pmatrix} 0 & 0 \\ 0 & Y \end{pmatrix} \:.
\eeq
We now vary~$M$ according to
\[ \tilde{M}_\tau = M + \tau\,e^{i \varphi} \begin{pmatrix} 0 & D \\ D^\dagger & 0 \end{pmatrix}
+ \frac{\tau^2}{2} \begin{pmatrix} X^{-1}\, E & 0 \\ 0 & 0 \end{pmatrix} \:. \]
As a consequence,
\begin{align*}
\tilde{M}_\tau^\dagger \,\tilde{M}_\tau = 
\begin{pmatrix} X^2 & 0 \\0 & 0 \end{pmatrix} + \tau \begin{pmatrix} 0 & e^{i \varphi} X D \\
e^{-i \varphi} D^\dagger X & 0 \end{pmatrix}
+ \tau^2\: \begin{pmatrix} D D^\dagger+E & 0 \\ 0 & D^\dagger D \end{pmatrix} \:.
\end{align*}
Under this variation, the constraints~\eqref{newconstr} behave as follows,
\begin{align*}
\Tr \big( S\, \tilde{M}_\tau^\dagger \,\tilde{M}_\tau \big) &= a + 2 \tau\, \re \Tr \bigg\{ e^{i \varphi} \:
S  \begin{pmatrix} 0 & X D \\ 0 & 0 \end{pmatrix} \bigg\}
+ \tau^2 \,\Tr \bigg\{ S \begin{pmatrix} D D^\dagger+E & 0 \\ 0 & D^\dagger D \end{pmatrix} \bigg\} \\
\Tr \big( \tilde{M}_\tau^\dagger \,\tilde{M}_\tau \big) &= b + 
\tau^2 \,\Tr \begin{pmatrix} D D^\dagger+E & 0 \\ 0 & D^\dagger D \end{pmatrix} \:.
\end{align*}
We now choose the phase~$\varphi$ such that the linear term in~$\tau$ vanishes.
Moreover, we choose~$E$ in such a way that the quadratic contributions in~$\tau$ vanish
(this can be done in all cases as explained in the proof of Lemma~\ref{lemmaL1}
after~\eqref{fvc}).

We have thus arranged that the variation satisfies the constraints.
Computing the second variation of~\eqref{auxMM} gives
\begin{align*}
0 &\geq \frac{d}{d\tau^2} \Tr \big(q \,S\, M^\dagger M \big) \big|_{\tau=0}
= \frac{d}{d\tau^2} \Tr \big(N\, M^\dagger M \big) \big|_{\tau=0} \\
&= 2\,\Tr \bigg\{ N\, \begin{pmatrix} D D^\dagger+E & 0 \\ 0 & D^\dagger D \end{pmatrix} \bigg\}
=  2\,\Tr \big\{ Y \,D^\dagger D \big\} \:,
\end{align*}
where in the last step we employed~\eqref{Nform}. Since~$D$ is arbitrary, we conclude that~$Y$
and therefore also~$N$ are positive semi-definite.
\QED

The result of this lemma can be understood in two ways, as a statement either on symmetric operators
on the indefinite inner product space~$(V, \Sl .|. \Sr)$ or on Hermitian matrices on~$\C^{2n}$.
We explain these different points of view after each other.
A positive semi-definite operator on an indefinite inner product space has a real spectrum
(see Lemma~\ref{lemmapositive}).
Moreover, the negative eigenvalues correspond to negative definite eigenspaces, whereas the
positive eigenvalues correspond to positive definite eigenspaces
(see again Lemma~\ref{lemmapositive} or~\cite{GLR}).
Interpreting the parameter~$\alpha$ as the eigenvalue of the operator~$q- \beta S$, we obtain the
following result.

\begin{Lemma} \label{lemmaalphaeigen}
The following statements hold:
\bitem
\item[{\rm{(i)}}] There is~$\alpha_0 \in \R$ such that all eigenvalues of the operator~$q-\beta S$
which are strictly smaller (strictly larger) than~$\alpha_0$ correspond to a negative (positive) definite eigenspace.
\item[{\rm{(ii)}}] Denoting all the parameters~$\alpha_0$ which satisfy~{\rm{(i)}} by~$\mathfrak{A}$, the
parameter~$\alpha$ in~\eqref{Arel} is on its boundary, i.e.\
\[ \alpha = \min \mathfrak{A} \qquad \text{or} \qquad \alpha = \max \mathfrak{A} \:. \]
\eitem
\end{Lemma}
\Proof This is an immediate consequence of Lemma~\ref{lemmapositive}.
\QED

The alternative point of view is to multiply the operator~$q-\alpha \1-\beta S$ from the left by~$S$, giving rise to the
positive semi-definite Hermitian matrix
\beq \label{positive}
\hat{q} - \alpha\,\1 S - \beta\,\1 \geq 0 \qquad \text{with} \qquad \hat{q} := S q \:.
\eeq
Moreover, we rewrite~\eqref{Arel} as
\beq \label{kernel}
\hat{A}\,\big(\hat{q}  - \alpha\, S - \beta\, \1 \big) = 0 \qquad \text{with} \qquad \hat{A}:= A S
\eeq
(the convention of multiplying by~$S$ from the left respectively right has the advantage
that~$Aq = \hat{A} \hat{q}$).
Interpreting~$\beta$ as the spectral parameter, by combining~\eqref{positive} with~\eqref{kernel}
one sees that~$\beta$ is the smallest eigenvalue of the matrix~$\hat{q} - \alpha S$,
\beq \label{betaval}
\beta(\alpha) = \min \sigma \big(\hat{q} - \alpha S \big) \:.
\eeq
This point of view also makes it possible to compute~$\hat{A}$ and to determine the
parameters~$a$ and~$b$ in~\eqref{auxconstr} as follows.

\begin{Prp} \label{prpunique} Choosing
\beq \label{bfix}
b = \Tr(\hat{A}) = 1 \:,
\eeq
the parameter~$a$ lies in the range
\[ a(\alpha) := \Tr(S \hat{A}) \in [-1,1] \:. \]
This function is monotone increasing and strictly monotone increasing except at the boundary points,
i.e.\
\[ \alpha_1 < \alpha_2 \quad \text{and} \quad |a(\alpha_1)|, |a(\alpha_2)|<1
\qquad \Longrightarrow \qquad a(\alpha_1) < a(\alpha_2) \:. \]
\end{Prp}
\Proof We form a spectral decomposition of the Hermitian matrix~$\hat{q}-\alpha S$,
\beq \label{specdec}
\hat{q}-\alpha S = \sum_{i=1}^{N} \lambda_j F_j \:,
\eeq
where~$F_j$ are spectral projection operators corresponding to the eigenvalues~$\lambda_j$,
which we label the eigenvalues in increasing order, i.e.\
\[ \beta = \lambda_1 < \lambda_2 < \cdots < \lambda_N \]
and~$N \leq 2n$.
We begin with the case that the smallest eigenvalue is non-degenerate.
In this case, the kernel of~$\hat{q}-\alpha S - \beta \1$ is one-dimensional, and the operator~$\hat{A}$
coincides, up to a prefactor, with the spectral projection operator~$F_1$.
The normalization condition~\eqref{bfix} fixes the prefactor to be one.
Hence, expressing the spectral projection with contour integrals (see for example~\cite{kato}), we obtain
\beq \label{hAval}
\hat{A} = F_1 = -\frac{1}{2 \pi i} \ointctrclockwise_\Gamma \big( \hat{q} - \alpha S - \lambda \big)^{-1}\: d\lambda \:,
\eeq
where the contour~$\Gamma$ encloses only the smallest eigenvalue with winding number one.
Multiplying by~$S$ and taking the trace, we obtain
(the reader not familiar with the perturbation theory for linear operators via contour integrals
may find it helpful to study~\cite[Appendix~G.1]{pfp})
\beq \label{aval}
a(\alpha) = \Tr(S \hat{A}) = -\frac{1}{2 \pi i} \ointctrclockwise_\Gamma \Tr \Big(
S \,\big( \hat{q} - \alpha\, S - \lambda\,\1 \big)^{-1} \Big)\: d\lambda \:.
\eeq
Differentiating with respect to~$\alpha$ gives
\[ a'(\alpha) = -\frac{1}{2 \pi i} \ointctrclockwise_\Gamma \Tr \Big( S\, \big( \hat{q} - \alpha\, S - \lambda \big)^{-1}
\,S\, \big( \hat{q} - \alpha\, S - \lambda \big)^{-1} \Big)\: d\lambda \:. \]
Plugging in a spectral decomposition~\eqref{specdec},
the contour integral can be computed with residues to obtain
\beq \label{alphap}
a'(\alpha) = \sum_{j=2}^{N} \frac{2}{\lambda_j-\lambda_1} \Tr \Big( S\, F_1
\,S\, F_j  \Big)\: d\lambda \:.
\eeq
For each summand, the term~$\lambda_j-\lambda_1$ is strictly positive. Moreover,
the operator~$S F_j S$ is a projection operator and thus positive, implying that the
trace in~\eqref{alphap} is non-negative. We conclude that~$a'(\alpha) \geq 0$.

In order to prove the strict inequality, let us assume that~$\alpha'(\alpha)=0$.
Then each summand in~\eqref{alphap} vanishes, implying that
\[ F_1 \, S \, F_j = 0 \qquad \text{for all~$j=2,\ldots, 2n$} \:. \]
As a consequence, $S$ maps the image of~$F_1$ to itself. In other words, the image of~$\hat{A}$
is an eigenspace of~$S$. This implies that~$a = \Tr(S \hat{A}) = \pm \Tr(\hat{A}) = \pm 1$.
This concludes the proof in the case that the lowest eigenvalue of~$\hat{q}-\alpha S$ is non-degenerate.

Now assume that~$\hat{q} - \alpha_0 S$ is degenerate. In this case, we can use perturbation theory
with degeneracies. In the first step, we need to analyze the perturbation operator~$S$
on the degenerate subspace, i.e.\ the operator
\[ F_1 S F_1 \::\: F_1(V) \rightarrow F_1(V)\:. \]
If this operator is a multiple of the identity, then we can use the perturbation theory without
degeneracies, and~\eqref{alphap} remains valid. Otherwise, for~$\alpha \neq \alpha_0$
and~$\alpha$ near~$\alpha_0$, the degeneracy of the lowest eigenvalue is removed.
The eigenspaces coincide to first order in~$\alpha-\alpha_0$ with those of the
operator~$-\epsilon(\alpha-\alpha_0) \,F_1 S F_1$ (where~$\epsilon$ is the sign function).
Consequently, if~$\hat{A}(\alpha)$ is a family of symmetric operators of trace one with
\[ \big( \hat{q} - \alpha\, S - \beta\, \1 \big) \hat{A}(\alpha) = 0 \:, \]
then the functional~$\Tr(S \hat{A})$ is discontinuous and monotone increasing at~$\alpha_0$
in the sense that
\[ \limsup_{\alpha \nearrow \alpha_0} \Tr \big(S \hat{A}(\alpha) \big)
= \min \sigma (F_1 S F_1) \big|_{\alpha_0} < \max \sigma (F_1 S F_1) \big|_{\alpha_0}
=\liminf_{\alpha \searrow \alpha_0} \Tr \big(S \hat{A}(\alpha) \big) \:. \]
This concludes the proof.
\QED

We point out that this proposition makes a general statement on how the smallest eigenvalue
of the matrix~$\hat{q}-\alpha S$ depends on~$\alpha$. It applies independent of the
context of the pointwise variational principle. In particular, it can be used to prove
the uniqueness statement of Proposition~\ref{prp91}:

\Proof[Proof of Proposition~\ref{prp91}]
In view of the inequality~\eqref{strictpoint}, we are in the case~$|a|<b$.
Therefore, by Proposition~\ref{prpunique}, the function~$a(\alpha)$ is strictly monotone increasing.
Consequently, there is at most one~$\alpha$ with~$a(\alpha)=a$.
For this value of~$\alpha$, the parameter~$\beta$ is uniquely determined by~\eqref{betaval}.
\QED

\subsection{Illustrating Examples}
We conclude this section with two simple examples.
The first example deals with the typical smooth situation when no degeneracies occur.

\begin{Example} {\em{ We consider the case~$n=1$ and choose
\[ S = \begin{pmatrix} 1 & 0 \\ 0 & -1 \end{pmatrix} \qquad \text{and} \qquad
q = i \sigma^2 = \begin{pmatrix} 0 & 1 \\ -1 & 0 \end{pmatrix} \]
(where~$\sigma^2$ is the second Pauli matrix).
This is a simple example intended to illustrate the case where the operators~$q$ and~$S$ do not commute
and the Lagrange parameters depend smoothly on~$a$ and~$b$.
Given~$a$ and~$b$ in the range~\eqref{abrange}, every symmetric operator satisfying the
constraints~\eqref{auxconstr} can be written as
\beq \label{Aansatz}
A = \frac{1}{2} \begin{pmatrix} a+b & -z \\ \overline{z} & a-b \end{pmatrix}
\eeq
with~$z \in \C$. In order for this matrix to be positive semi-definite, the determinant of~$AS$ must be
positive, meaning that
\beq \label{zrange}
|z|^2 \leq b^2-a^2 \:.
\eeq
The functional in the pointwise variational principle is computed by~$\Tr(q A) = \re z$.
Minimizing this functional in the region~\eqref{zrange} gives a unique minimizer at~$z = -\sqrt{ b^2-a^2 }$.
Hence the minimizer~$A$ takes the form
\beq \label{Akernel}
A = \frac{1}{2} \begin{pmatrix} a+b & \sqrt{b^2-a^2} \\[0.2em] -\sqrt{b^2-a^2} & a-b \end{pmatrix} \:.
\eeq

The matrix~$q  - \alpha \1- \beta S$ in the statement of Lemma~\ref{lemmaL1} takes the form
\[ q  - \alpha\, \1 - \beta\, S = \begin{pmatrix} -\alpha-\beta & 1 \\ -1 & -\alpha+\beta \end{pmatrix} \:. \]
The image of this matrix must be in the kernel of the matrix~$A$ in~\eqref{Akernel}. This is the case
if and only if
\beq \label{albe}
\alpha = \frac{a}{\sqrt{b^2-a^2}} \qquad \text{and} \qquad \beta = -\frac{b}{\sqrt{b^2-a^2}} \:,
\eeq
giving
\[ q  - \alpha\, \1 - \beta\, S = \frac{1}{\sqrt{b^2-a^2}}
\begin{pmatrix} -a+b & \sqrt{b^2-a^2} \\ -\sqrt{b^2-a^2} & -a-b \end{pmatrix} \:. \]
This matrix is indeed positive semi-definite.

In order to see the connection to the statement of Lemma~\ref{lemmaalphaeigen}, we first note
that the operator~$q-\beta S$ has the eigenvalues
\[\alpha_\pm = \pm \sqrt{\beta^2-1} \:. \]
In order for the operator~$q-\beta S$ to be positive, we need to choose~$\beta < -2$,
in agreement with the right equation in~\eqref{albe}. The set~$\mathfrak{A}$ is computed by
\[ \mathfrak{A} = \big[ -\sqrt{\beta^2-1}, \sqrt{\beta^2-1} \big] \:. \]
The two boundary points of this interval give us back the values for~$\alpha$ in~\eqref{albe}
for positive respectively negative~$a$.

In order to get into the setting of Proposition~\ref{prpunique}, we compute the matrix~$\hat{q}-\alpha S$,
\[ \hat{q}-\alpha S = \begin{pmatrix} -\alpha & 1 \\ 1 & \alpha \end{pmatrix} \:. \]
It has the spectral decomposition~\eqref{specdec} with eigenvalues
\[ \lambda_1 = -\sqrt{1+\alpha^2}\:,\qquad \lambda_2 = \sqrt{1+\alpha^2} \]
and spectral projection operators
\[ F_{1\!/\!2} = \frac{1}{2 \,\sqrt{1+\alpha^2}}
\begin{pmatrix} \sqrt{1+\alpha^2} \pm \alpha & \mp 1 \\ \mp 1 &  \sqrt{1+\alpha^2} \mp \alpha \end{pmatrix}\:. \]
Using~\eqref{betaval}, \eqref{hAval} and~\eqref{aval}, 
we obtain
\begin{align*}
\beta &=\lambda_1 = -\sqrt{1+\alpha^2} \\
\hat{A} &= F_1 = \frac{1}{2 \,\sqrt{1+\alpha^2}}
\begin{pmatrix} \sqrt{1+\alpha^2} + \alpha & -1 \\ -1 &  \sqrt{1+\alpha^2} - \alpha \end{pmatrix} \\
a &= \Tr(S \hat{A}) = \frac{\alpha}{\sqrt{1+\alpha^2}} \:.
\end{align*}
These formulas agree with~\eqref{Akernel} and~\eqref{albe} if~$\alpha$ and~$\beta$ are expressed in
terms of~$a$ and~$b$.

Our findings are illustrated in Figure~\ref{figex2}.
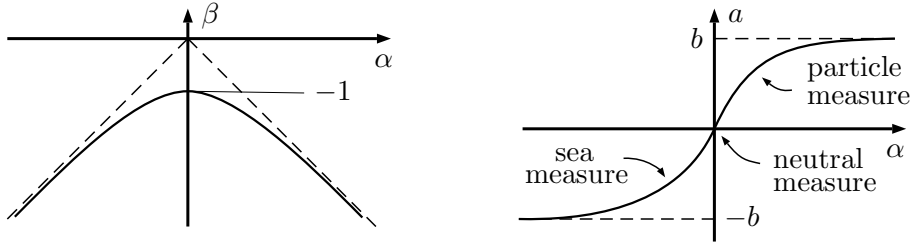
\begin{figure}
\psscalebox{1.0 1.0} 
{
\begin{pspicture}(0,26.864866)(12.006803,30.014597)
\psline[linecolor=black, linewidth=0.04, arrowsize=0.05291667cm 2.0,arrowlength=1.4,arrowinset=0.0]{->}(0.017070312,29.574865)(5.11707,29.574865)
\psline[linecolor=black, linewidth=0.04, arrowsize=0.05291667cm 2.0,arrowlength=1.4,arrowinset=0.0]{->}(6.86707,28.374866)(11.967071,28.374866)
\psline[linecolor=black, linewidth=0.04, arrowsize=0.05291667cm 2.0,arrowlength=1.4,arrowinset=0.0]{->}(2.4170704,27.074865)(2.4170704,29.974865)
\psline[linecolor=black, linewidth=0.04, arrowsize=0.05291667cm 2.0,arrowlength=1.4,arrowinset=0.0]{->}(9.41707,26.874866)(9.41707,29.974865)
\psline[linecolor=black, linewidth=0.02, linestyle=dashed, dash=0.17638889cm 0.10583334cm](2.4170704,29.574865)(4.9170704,27.074865)
\psline[linecolor=black, linewidth=0.02, linestyle=dashed, dash=0.17638889cm 0.10583334cm](2.4170704,29.574865)(0.017070312,27.174866)
\psline[linecolor=black, linewidth=0.02, linestyle=dashed, dash=0.17638889cm 0.10583334cm](6.8170705,27.174866)(9.41707,27.174866)
\psline[linecolor=black, linewidth=0.02, linestyle=dashed, dash=0.17638889cm 0.10583334cm](9.41707,29.574865)(11.81707,29.574865)
\psbezier[linecolor=black, linewidth=0.03](0.11707031,27.204866)(0.9516386,28.050909)(1.7970703,28.874866)(2.4170704,28.87486572265625)(3.0370703,28.874866)(3.972502,27.97091)(4.73707,27.194866)
\psbezier[linecolor=black, linewidth=0.03](6.8170705,27.174866)(7.632643,27.157652)(8.91707,27.274866)(9.41707,28.37486572265625)(9.91707,29.474865)(10.485029,29.543924)(11.81707,29.574865)
\psline[linecolor=black, linewidth=0.01](2.4170704,28.874866)(3.9632242,28.85179)
\psbezier[linecolor=black, linewidth=0.02, arrowsize=0.05291667cm 2.0,arrowlength=1.4,arrowinset=0.0]{->}(10.44784,28.890251)(10.378609,28.882559)(10.278608,28.867174)(10.047839,29.059481107271576)
\psbezier[linecolor=black, linewidth=0.02, arrowsize=0.05291667cm 2.0,arrowlength=1.4,arrowinset=0.0]{->}(10.024762,27.913328)(9.847839,27.951788)(9.763225,27.95948)(9.509378,28.28255803034852)
\psbezier[linecolor=black, linewidth=0.02, arrowsize=0.05291667cm 2.0,arrowlength=1.4,arrowinset=0.0]{->}(8.209378,27.982557)(8.540147,27.997942)(8.717071,27.797943)(8.763225,27.728711876502345)
\rput[bl](4.9,29.2){$\alpha$}
\rput[bl](2.6,29.7){$\beta$}
\rput[bl](4.1,28.73){$-1$}
\rput[bl](11.7,28){$\alpha$}
\rput[bl](9.6,29.8){$a$}
\rput[bl](9.1,29.45){$b$}
\rput[bl](9.55,27.05){$-b$}
\rput[bl](10.65,29){particle}
\rput[bl](10.65,28.7){measure}
\rput[bl](7.3,27.95){sea}
\rput[bl](6.9,27.65){measure}
\rput[bl](10.2,27.85){neutral}
\rput[bl](10.2,27.55){measure}
\end{pspicture}
}
\caption{The Lagrange multipliers in the example~$q=i \sigma^2$.}
\label{figex2}
\end{figure}%
We point out that, for all values of~$\alpha$,
the operator~$q-\alpha \1-\beta S$ has a one-dimensional kernel. Correspondingly, the operator~$A$
has rank one. Its non-zero eigenvalue coincides with its trace~$a$. Therefore, one can immediately
read off the decomposition of Definition~\ref{defnudecomp}: For negative~$a$, one has a sea
measure, for positive~$a$ a particle measure, whereas the intermediate case~$a=0$
gives a neutral measure.
}} \QEDrem
\end{Example}

The next example illustrates the case with degeneracies.
\begin{Example} {\em{ We consider the case~$n=1$ and
\[ q = S = \begin{pmatrix} 1 & 0 \\ 0 & -1 \end{pmatrix} \:. \]
Given~$a$ and~$b$ in the range~\eqref{abrange}, the symmetric matrices
satisfying the constraints~\eqref{auxconstr} are again of the form~\eqref{Aansatz}
with~$z \in \C$ in the region~\eqref{zrange}.
Now the functional in the pointwise variational principle is computed by~$\Tr(q A) = b$.
This functional is determined by the constraints. Therefore, there is nothing to vary,
and every~$A$ of the form~\eqref{Aansatz} is a minimizer. This is consistent with
the statement of Lemma~\ref{lemmaL1}, leaving us some freedom to choose~$\alpha$ and~$\beta$.
One choice is
\beq \label{case0}
\alpha=0, \:\beta=1 \qquad \Longrightarrow \qquad q  - \alpha\, \1 - \beta\, S = 0 \:,
\eeq
giving us the freedom to choose~$A$ arbitrarily according to~\eqref{Aansatz}.
Alternatively, one can choose
\beq \label{case1}
\alpha<0, \:\beta=1+\alpha \qquad \Longrightarrow \qquad q  - \alpha \1 - \beta\, S =
\begin{pmatrix} -2 \alpha & 0 \\ 0 & 0 \end{pmatrix}  \:,
\eeq
in which case in~\eqref{Aansatz} we must choose~$a=-b$ and~$z=0$.
Finally, one can also choose
\beq \label{case2}
\alpha>0, \:\beta=1-\alpha \qquad \Longrightarrow \qquad q  - \alpha \1 - \beta\, S =
\begin{pmatrix} 0 & 0 \\ 0 & -2 \alpha \end{pmatrix}  \:,
\eeq
in which case in~\eqref{Aansatz} we must choose~$a=b$ and~$z=0$.
Note that in~\eqref{case1} and~\eqref{case2}, the sign of~$\alpha$ is determined by
the requirement that the matrix~$q  - \alpha \1 - \beta S$ be positive semi-definite
with respect to~$\Sl .|. \Sr$.

These findings fit together with the statement of Lemma~\ref{lemmaalphaeigen} as follows.
The operator~$q-\beta S$ has the eigenvalues~$\pm (1 -\beta)$. In order for the smaller (larger)
eigenvalue to correspond to a negative (positive) definite eigenspace, we need to choose~$\beta \leq 1$.
In this case, the set~$\mathfrak{A}$ is given by
\[ \mathfrak{A} = \big[ -(1-\beta), 1-\beta \big] \:, \]
giving us the possible choices~$\alpha= \pm (1-\beta)$ in~\eqref{case1} and~\eqref{case2}.

In the setting of Proposition~\ref{prpunique}, the matrix~$\hat{q}$ is the identity.
In the spectral decomposition of~$\hat{q}-\alpha S$ in~\eqref{specdec} we need to distinguish the
following cases:
\bitem
\itemD $\alpha=0$: In this case, $N=1$ and
\[ \lambda_1 = 1\:,\qquad F_1 = \1 \:. \]
In this case, $A$ can be any operator of the form~\eqref{Aansatz}, giving back case~\eqref{case0} above.
\itemD $\alpha<0$: In this case, $N=2$ and
\[ \beta = \lambda_1 = 1+\alpha\:,\;\;\; F_1 = \begin{pmatrix} 0 & 0 \\ 0 & 1 \end{pmatrix} \qquad \text{and} \qquad
\lambda_2 = 1-\alpha\:,\;\;\; F_2 = \begin{pmatrix} 1 & 0 \\ 0 & 0 \end{pmatrix} \:. \]
In order for~$A$ to be a multiple of~$F_1$, we need to choose~$a=-b$ and~$z=0$.
This corresponds precisely to case~\eqref{case1} above.
\itemD $\alpha>0$: In this case, $N=2$ and
\[ \beta =\lambda_1 = 1-\alpha\:,\;\;\; F_1 = \begin{pmatrix} 1 & 0 \\ 0 & 0 \end{pmatrix} \qquad \text{and} \qquad
\lambda_2 = 1+\alpha\:,\;\;\; F_2 = \begin{pmatrix} 0 & 0 \\ 0 & 1 \end{pmatrix} \:. \]
In order for~$A$ to be a multiple of~$F_1$, we need to choose~$a=b$ and~$z=0$.
This corresponds precisely to case~\eqref{case2} above.
\eitem
Our findings are illustrated in Figure~\ref{figex1}.
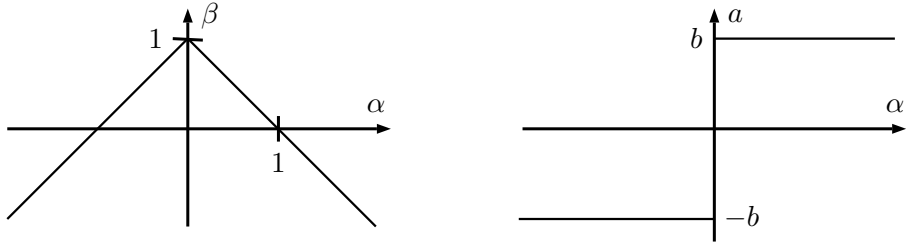
\begin{figure}
\psscalebox{1.0 1.0} 
{
\begin{pspicture}(0,26.864866)(12.010338,30.014597)
\psline[linecolor=black, linewidth=0.04, arrowsize=0.05291667cm 2.0,arrowlength=1.4,arrowinset=0.0]{->}(0.02060547,28.374866)(5.1206055,28.374866)
\psline[linecolor=black, linewidth=0.04, arrowsize=0.05291667cm 2.0,arrowlength=1.4,arrowinset=0.0]{->}(6.8706055,28.374866)(11.970606,28.374866)
\psline[linecolor=black, linewidth=0.04, arrowsize=0.05291667cm 2.0,arrowlength=1.4,arrowinset=0.0]{->}(2.4206054,27.074865)(2.4206054,29.974865)
\psline[linecolor=black, linewidth=0.04, arrowsize=0.05291667cm 2.0,arrowlength=1.4,arrowinset=0.0]{->}(9.420606,26.874866)(9.420606,29.974865)
\psline[linecolor=black, linewidth=0.03](2.4206054,29.574865)(4.9206057,27.074865)
\psline[linecolor=black, linewidth=0.03](2.4206054,29.574865)(0.02060547,27.174866)
\psline[linecolor=black, linewidth=0.03](6.8206053,27.174866)(9.420606,27.174866)
\psline[linecolor=black, linewidth=0.03](9.420606,29.574865)(11.820605,29.574865)
\psline[linecolor=black, linewidth=0.04](2.2206054,29.574865)(2.6206055,29.554865)
\psline[linecolor=black, linewidth=0.04](3.6256056,28.544867)(3.6256056,28.204866)
\rput[bl](4.8,28.6){$\alpha$}
\rput[bl](2.6,29.7){$\beta$}
\rput[bl](11.7,28.6){$\alpha$}
\rput[bl](9.6,29.8){$a$}
\rput[bl](9.1,29.45){$b$}
\rput[bl](9.55,27.05){$-b$}
\rput[bl](1.9,29.45){$1$}
\rput[bl](3.53,27.8){$1$}
\end{pspicture}
}
\caption{The Lagrange multipliers in the example~$q=S$.}
\label{figex1}
\end{figure}%
If~$\alpha$ is non-zero, the operator~$q-\alpha \1 -\beta S$ has a one-dimensional kernel,
and the operator~$A$ has rank one. From the sign of its trace we can again read off
the decomposition of Definition~\ref{defnudecomp}: For negative~$\alpha$, one has a sea
measure and for positive~$\alpha$ a particle measure.
The fact that the function~$a(\alpha)$ is locally constant means that, given~$a$ and~$b$
with~$|a|=b$, the Lagrange parameters~$\alpha$ and~$\beta$ are not unique.
In the case~$\alpha=0$ and~$\beta=1$, the operator~$q-\alpha \1-\beta S$ vanishes.
As a consequence, the operator~$A$ can be chosen arbitrarily according to~\eqref{Aansatz}.
This means that the choice of Lagrange parameters~$\alpha=0$ and~$\beta=1$ is admissible for
any~$a \in [-b,b]$.
}} \QEDrem
\end{Example}

\section{Discussion and Outlook} \label{secoutlook}
In this paper we gave a detailed analysis of the homogeneous causal action principle
on a compact domain~$\hat{K}$ of momentum space.
The derived EL equations have an interesting mathematical structure.
In order to put this result into context, we remark that this structure has some similarity
to the notion of {\em{state stability}} introduced in~\cite[Section~5.6]{pfp}.
Indeed, if one restricts attention to sea measures (see Definition~\ref{defnudecomp}),
specifies to a vector-scalar structure of~$\hat{Q}$ 
and leaves out the dimension constraint, then the minimality statement in Theorem~\ref{thmELalt}
goes over to the minimality statement in~\cite[Definition~5.6.2~(iii)]{pfp}.
Therefore, our results confirm the considerations on the stability of the Minkowski vacuum
in~\cite[Section~5.6]{pfp} and put these considerations on a solid mathematical basis.

It is unknown whether the homogeneous causal action principle remains well-posed
if one {\em{removes the compactness assumption}} on~$\hat{K}$. In particular, it is an important
problem to understand the behavior in the case~$\hat{K}=\R^4$ where the measure~$\nu$
can be supported anywhere in Minkowski space. It is not clear whether this variational principle
is well-posed. The difficulties can be understood in analogy to the ultraviolet problems in quantum
field theory: Choosing a compact domain~$\hat{K}$ can be understood as introducing an a-priori
momentum cutoff. If this cutoff is removed,
the homogeneous variational principle might well develop singularities in analogy to the
divergences in quantum field theory. In this case, inspired by the renormalization procedure
in quantum field theory, the strategy would be to take the limit~$\hat{K} \nearrow \R^4$
after suitably rescaling the Lagrangian. We note for clarity that these divergences are
a consequence that the homogeneous causal action principle typically involves an infinite
number of particles. In contrast, the causal action principle in~\cite[Section~1.1]{cfs}
formulated on a finite-dimensional Hilbert space is well-posed and finite.
With this in mind, the non-homogeneous causal action principle should be considered as being more fundamental,
whereas the homogeneous setting merely is an approximation valid on certain energy scales.

We finally explain how the homogeneous causal action principle on the non-compact domain~$\hat{K}=\R^4$
could be attacked starting from the methods and results presented here.
Choosing an exhaustion~$\hat{K}_n$ of~$\R^4$,
\[ \hat{K}_1 \subset \hat{K}_2 \subset \cdots \qquad \text{and} \qquad \bigcup_{\ell=1}^\infty
\hat{K}_\ell=\R^4\:, \]
one considers a sequence~$\nu_\ell$ of minimizing measures on~$\hat{K}_\ell$.
Then each measure~$\nu_\ell$ satisfies the EL equations on~$\hat{K}_\ell$.
The hope is that a subsequence of these measures converges in a suitable topology
to a measure~$\nu$ which satisfies the EL equations in all of~$\R^4$.
This strategy was already implemented in position space in~\cite{noncompact}.
However, making it work in momentum space is more challenging for several reasons.
One difficulty is that the EL equations of the homogeneous causal action principle
(as stated for example in Theorem~\ref{thmEL}) are more involved and seem to give
less control of the measure~$\nu$. One should also keep in mind that these EL equations
require suitable regularity assumptions (in particular, the Lagrangian must be differentiable
in the sense~\eqref{delLdef}), which might make it necessary to ``regularize''
the Lagrangian (i.e.\ to replace the causal Lagrangian~$\L$ by a smooth Lagrangian~$\L_\ell$
which tends to~$\L$ in the limit~$\ell \rightarrow \infty$).
Next, the homogeneous causal action principle involves
a larger rescaling freedom, which cannot be fixed in an obvious way. One freedom is to
translate the measure by a vector~$\Lambda \in \R^4$ by setting
\beq \label{nugauge}
\nu_\Lambda(\Omega) := \nu(\Omega-\Lambda) \:.
\eeq
Changing variables in~\eqref{Preg}, one sees that this translation in momentum space
merely gives rise to a phase factor~$e^{-i\Lambda \xi}$, which drops out of the Lagrangian.
This phase factor can be regarded as a gauge phase, and the freedom~\eqref{nugauge}
can be understood as a {\em{gauge freedom}}. Using the language of gauge theory,
it is not clear how to fix the gauge in a canonical way.
Apart from this gauge freedom, 
one can also transform the measure by a linear transformation~$A \in \Lin(\R^4)$,
\beq \label{nulinear}
\nu_A(\Omega) := \nu \big( A^{-1} \,\Omega \big) \:.
\eeq
Such a linear transformation maps a minimizer~$\hat{K}_\ell$ to a minimizer on~$A \hat{K}_\ell$,
with the parameters~$c$ and~$f$ of the constraints unchanged.
Finally, one can also rescale the measure by a factor~$\lambda>0$,
\[ \nu_\lambda := \lambda \nu \:. \]
This rescaling again maps minimizers to minimizers, with the constraints linearly scaled,
\[ c_\lambda = \lambda c \qquad \text{and} \qquad f_\lambda = \lambda f \:. \]
The hope is that, after suitably applying the above transformations to all the~$\nu_\ell$
and possibly after renormalizing the causal Lagrangian,
one gets the desired convergence to a solution~$\nu$ of the EL equations.

\Thanks{{{\em{Acknowledgments:}}
We would like to thank the referee for helpful comments on the manuscript.

\providecommand{\bysame}{\leavevmode\hbox to3em{\hrulefill}\thinspace}
\providecommand{\MR}{\relax\ifhmode\unskip\space\fi MR }
\providecommand{\MRhref}[2]{%
  \href{http://www.ams.org/mathscinet-getitem?mr=#1}{#2}
}
\providecommand{\href}[2]{#2}

\end{document}